\documentclass[11pt,dvips]{article}
\usepackage{amscd,verbatim}
\usepackage[all]{xy}
\usepackage{graphicx}

\usepackage{amsmath}
\usepackage[psamsfonts]{amssymb}
\usepackage{amsthm}
\usepackage{euscript}

\usepackage{latexsym}

\renewcommand{\(}{\begin{equation}}
\renewcommand{\)}{end{equation} \vspace{-.05in}\linebreak}

\newcounter{saveeqn}
\newcounter{savealpheqn}

\newcommand{\alpheqn}{\setcounter{saveeqn}{\value{equation}}%
 \stepcounter{saveeqn}\setcounter{equation}{0}%
 \renewcommand{\theequation}{\mbox{\arabic{section}.\arabic{saveeqn}
\alph{equation}}}
 \renewcommand{\)}{\end{equation}}}
\def\part#1{\frac{\partial}{\partial{#1}}}%
\def\group#1{\refstepcounter{equation}\setcounter{saveeqn}{\value{equation}}%
 \label{#1}\setcounter{equation}{0}%
\renewcommand{\theequation}{\mbox{\arabic{section}.\arabic{saveeqn}
\alph{equation}}}
 \renewcommand{\)}{\end{equation}}}
\newcommand{\reseteqn}{\setcounter{equation}{\value{saveeqn}}%
 \renewcommand{\theequation}{\arabic{section}.\arabic{equation}}%
 \renewcommand{\)}{\end{equation}}}

\newcommand{\aalpheqn}{\setcounter{saveeqn}{\value{equation}}%
 \stepcounter{saveeqn}\setcounter{equation}{0}%
 \renewcommand{\theequation}{\mbox{
       \Alph{subsection}.\arabic{saveeqn}\alph{equation}}}
  \renewcommand{\)}{\end{equation}}}
\newcommand{\areseteqn}{\setcounter{equation}{\value{saveeqn}}%
 \renewcommand{\theequation}{\Alph{subsection}.\arabic{equation}}%
 \renewcommand{\)}{\end{equation}}}

\renewcommand{\thefootnote}{\alph{footnote}}
\renewcommand{\(}{\begin{equation}}
\renewcommand{\)}{\end{equation}}
\newcommand{\ba}{\begin{eqnarray}}
\newcommand{\ea}{\end{eqnarray}}

\newcommand{\bp}{\mathop{\vtop{\ialign{##\crcr
  $\hfil\displaystyle{}\hfil$\crcr\noalign{\kern-13pt\nointerlineskip}
  \BIG{(}\hskip0pt\crcr\noalign{\kern3pt}}}}}
\newcommand{\cbp}{\mathop{\vtop{\ialign{##\crcr
  $\hfil\displaystyle{}\hfil$\crcr\noalign{\kern-13pt\nointerlineskip}
  \BIG{)}\hskip0pt\crcr\noalign{\kern3pt}}}}}
\newcommand{\pa}{\mathop{\vtop{\ialign{##\crcr
  $\hfil\displaystyle{\oplus}\hfil$\crcr\noalign{\kern+1pt\nointerlineskip}
  \hspace{.08in}$^{\alpha=0}$\hskip6pt\crcr\noalign{\kern3pt}}}}}
\renewcommand{\sp}{,\hspace{.3in}}
\newcommand{\p}{^\prime}

\newcommand{\R}{\ensuremath{\mathbb R}}

\newcommand{\C}{\ensuremath{\mathbb C}}

\newcommand{\Z}{\ensuremath{\mathbb Z}}

\newcommand{\beq}{\begin{equation}}
\newcommand{\eeq}{\end{equation}}

\numberwithin{equation}{section}

\def\hsp#1{\hspace{#1in}}

\catcode`\@=11
\def\vereq#1#2{\lower3pt\vbox{\baselineskip1.5pt \lineskip1.5pt
\ialign{$\m@th#1\hfill##\hfil$\crcr#2\crcr\sim\crcr}}}
\catcode`\@=12

\makeatletter
\newcommand\figcaption{\def\@captype{figure}\caption}
\newcommand\tabcaption{\def\@captype{table}\caption}
\makeatother
\renewcommand{\(}{\begin{equation}}
\renewcommand{\)}{\end{equation}}

\newcommand{\RR}{{\mathbb R}}

\theoremstyle{plain}

\theoremstyle{definition}

\newcommand{\twoa}{\text{I}\!\text{IA}}
\newcommand{\twob}{\text{I}\!\text{IB}}

\begin{document}

\begin{titlepage}
\begin{flushright}
ULB-TH-04/18

hep-th/0405210
\end{flushright}

\vspace{2em}
\def\thefootnote{\fnsymbol{footnote}}

\begin{center}
{\Large\bf The Cascade is a MMS Instanton}
\end{center}
\vspace{1em}

\begin{center}
Jarah Evslin\footnote{E-Mail: jarah@df.unipi.it}\ 
\end{center}

\begin{center}
\vspace{1em}
\hsp{.3}\\
{\small
Physique Th\'eorique et Math\'ematique,\\
Universit\'e Libre
de Bruxelles,\\C.P. 231, B-1050, Bruxelles, Belgium}

\end{center}

\vspace{8em}
\begin{abstract}
\noindent
Wrap $m$ D5-branes around the 2-cycle of a conifold, place $n$ D3-branes at a point and watch the system relax.  The D5-branes source $m$ units of RR 3-form flux on the 3-cycle, which cause dielectric NS5-branes to nucleate and repeatedly sweep out the 3-cycle, each time gaining $m$ units of D3-charge while the stack of D5-branes loses $m$ units of D3-charge.  A similar description of the Klebanov-Strassler cascade has been proposed by Kachru et al. when $m>>m-n$, where it is a tunneling event in the dual field theory.   Using the T-dual MQCD we argue that the above process occurs for any $m$\ and $n$ and in particular may continue for more than one step.  The nonbaryonic roots of the SQCD vacua lead to new cascades because, for example, the 3-cycle swept does not link all of the D5's.  This decay is the S-dual of a MMS instanton, which is the decay into flux of a brane that is trivial in twisted K-theory.  This provides the first evidence for the S-dual of the K-theory classification that does not itself rely upon any strong/weak duality.
\end{abstract}

\vfill

23 May, 2004 

\end{titlepage}
\setcounter{footnote}{0} 
\renewcommand{\thefootnote}{\arabic{footnote}}

\pagebreak
\renewcommand{\thepage}{\arabic{page}}

\section{Introduction}

The reader interested in the cascade but not the K-theory classification may skip to Subsec. ~\ref{mmssec}.

Given a compactification of type II string theory, how can we classify the topologically stable configurations?  If we are only interested in configurations without branes, then we need to classify fluxes, which are differential forms.  Classically this requires considering fluxes that solve the equations of motion modulo the applicable gauge invariances, quantum mechanically one may also need to apply a quantization condition and include some torsion corrections to the equations of motion.  Considering the equations of motion and gauge invariances
\begin{equation}
dG_{p+1}=H\wedge G_{p-1}\sp C_p\longrightarrow C_p+d\Lambda_{p-1}+H\wedge\Lambda_{p-3} \label{bianchi}
\end{equation}
of the Ramond-Ramond fluxes defined below one finds, as is reviewed in section~\ref{revsec}, that after including torsion corrections RR fluxes are classified by twisted K-theory \cite{MM,WKTheory,Manjarin}.

This construction realizes twisted K-theory as a quotient (gauge orbits) of a subset (solutions of equations of motion) of integral cohomology.  One may try to apply the same logic to conclude that D-branes are classified by twisted K-homology, which is a quotient of a subset of integral homology.  However in the case of D-branes the states that should be excluded in fact are legitimate solitons, they are the baryons of Ref.~\cite{Baryons}.  In addition the D-brane configurations that should be equivalent are not quite equivalent.  There are dynamical processes, called MMS instantons, that turn a configuration into another that represents the same K-homology class \cite{MMS}.  While successfully interpolating between homologically distinct D-branes, these processes also change the background fluxes.  The new flux can have dramatic physical effects.  For example it can make a type IIA theory massive \cite{MMS,Feb} or, as is the subject of this paper, it can be the $H$ flux produced during the Klebanov-Strassler \cite{KS} cascade and thus contribute to the superpotential of the corresponding $N=1$ gauge theory.  In the case of the D3-brane decay described in the abstract the resulting flux implies that for each D3-brane created by an NS5-brane a D5-brane loses a unit of D3-charge.

The twisted K-theory classification of the RR fluxes faces similar problems.  The fluxes corresponding to cohomology classes that are not K-theory classes are configurations with branes, and so rather than being unphysical, like the baryons they merely lie outside the set of configurations considered.  Meanwhile the field strengths that are trivial in twisted K-theory may be gauged away, but this transformation leaves a residual integral, globally defined gauge field, which is a higher dimensional analogue of a theta angle that is a multiple of $2\pi$.  While this may seem trivial, if we allow NS5-branes to sweep out cycles in our spacetime until $H$ integrated over every cycle vanishes, any flux that seemed to have been gauged away will be restored\footnote{This is a result of the Chern-Simons supergravity coupling of the theta angle to $H$ and the restored field strength.  In the quantum theory this process, if possible, must continue until $W_3+H=0$ on each cycle, and correspondingly the gauge transformations must be augmented by Steenrod square terms so that the same cancellation occurs.} \cite{Nov}.  Thus the theta angles remember the original flux configuration, and so the fluxes corresponding to trivial twisted K-theory classes cannot in fact be pure gauge.  

Even if we ignore these problems by considering only configurations without branes and identifying configurations with different theta angles, RR fluxes are not classified by twisted K-theory.  In addition to the equations of motion and Bianchi identities (\ref{bianchi}) for the RR fields which yield the K-theory classification, one must also impose the equations of motion and quotient by the Bianchi identity of the NS 3-form field strength $H=dB$ and its dual $*H=dB^{(6)}$.  In type IIB string theory, for example these are easily found by S-duality to be
\begin{equation}
d*H=G_3\wedge G_5\sp B\longrightarrow B+d\Lambda_1 \sp B^{(6)}\longrightarrow B^{(6)}+d\Lambda_5+G_3\wedge \Lambda_3 \label{nseom}
\end{equation}
where $\Lambda_3$ is the same 3-form used to simultaneously transform $C_4$. It is tempting to S-dualize $dG_3=H\wedge G_1$ to find $dH$, however when the right hand side is nontrivial the usual arguments for S-duality fail, and indeed one appears to find that the dilaton is not globally defined.  While such bizarre configurations do appear, for example, in F-theory on K3 \cite{Ftheory}, it seems impossible to disentangle them from the topology of the ambient spacetime and so they apparently defy the kind of classification scheme with which we are concerned here.   Luckily no such flux appears in the Klebanov-Strassler geometry.

The extra restrictions and quotients of Eq.~(\ref{nseom}) imply that RR fluxes are classified by something very different from twisted K-theory.  In fact, it has been known for some time \cite{DMW} that the K-theory classification is not consistent with S-duality in type IIB string theory.  Thus the classification of branes in type IIB must be some S-duality covariant modification of K-theory, where the above problems with the twisted K-theory classification are hidden by excluding configurations with baryons, identifying configurations with different fluxes and cutting out any NS5-branes that source these fluxes, or in the S-dual version of interest cutting out the D5-brane sources.  The usual construction of K-theory from vector bundles does not appear to admit an obvious covariantization.  A more promising construction, by Maldacena, Moore and Seiberg (MMS) in Ref.~\cite{MMS}, uses physical processes, called MMS instantons.  While they consider instantons constructed entirely from D-branes, the construction appears to work with fundamental strings and NS5-branes as well and so, modulo problems with 7-branes discussed above, is easily made S-duality covariant.  Using this S-duality covariantized MMS construction a proposed S-duality covariant classification of branes and strings in type IIB was proposed in Refs.~\cite{Uday,Nov}. 

The only nontrivial test of this proposal to date has been by Loaiza-Brito in Ref.~\cite{Oscar}, who has shown that it is consistent with the $SL(2,\Z)$ duality of 4-dimensional $N=4$ supersymmetric gauge theories.  In this note we provide a second test.  D3-branes at a point on the deformed conifold with $m$ units of $G_3$ flux are classified by $\Z$ in ordinary twisted K-theory, and so should never decay.  However in the S-duality covariant generalization, and even in the S-dual of the ordinary K-theory classification, they are classified by $\Z_m$, and so $m$ D3-branes may decay at a time.  The cascade is precisely a series of such decays, and so it is consistent with the S-duality covariant K-theory classification and not with the ordinary K-theory classification.  Furthermore the MMS instanton that corresponds to this decay is the decay described at large $m$ by Kachru, Pearson and Verlinde (KPV) in Ref.~\cite{KPV}, where our use of the T-dual MQCD eliminates the need to introduce auxilliary anti-branes to break the supersymmetry and so drive this process\footnote{The process is driven by the tension of the M5-brane, which is lost in the 10d $g_s=0$ dimensional reduction.}.  As described above, such instantons leave a residual flux.  In this case there is one unit of $H$ flux for every $m$ D3-branes that decay.  Using the T-dual MQCD we are able to extend these results to small $m$ and $n$, where the usual arguments for the deformation of the conifold do not apply.  We will find that D-branes are still classified by the (S-dual) twisted K-theory of the deformed conifold.

We begin in Sec.~\ref{revsec} with a review of the twisted K-theory classification of fluxes and branes.  We argue that Page field strengths correspond to the integral cohomology classes of the Atiyah-Hirzebruch spectral sequence while gauge-invariant field strengths are the twisted Chern characters \cite{BCMMS,MathStev} of the twisted K-theory classes.  In Sec.~\ref{wzwsec} we recall that the $SU(2)$ WZW model also exhibits a cascade and show that this cascade is U-dual to a process similar to that of KPV and valid at all $m$\ and $n$.  In Sec.~\ref{mqcdsec} we review the Klebanov-Strassler cascade in MQCD, where we can see features not readily apparent in the conifold description.  In addition to the baryonic vacua of the theory studied by Klebanov and Strassler \cite{KS}, one also sees the nonbaryonic vacua which lead to different cascades.  We consider generalizations of the 4d gauge theory, such as $N=2$ softly broken to $N=1$ by a superpotential that is a nondegenerate polynomial in the adjoint chiral multiplet, where one finds copies of the same story located at each extremum of the superpotential. Finally in Sec.~\ref{dualsec} we T-dualize this description to the conifold, where we find the KPV process.

\section{The Twisted K-Theory Classification} \label{revsec}

\subsection{Classifying Field Strengths}

If $C_p$ are the RR gauge fields and $B$ the NS two-form then there are at least two kinds of RR field strengths that we may be interested in classifying \cite{Marolf}.  The first is the gauge invariant
\begin{equation}
G_{p+1}=dC_p-H\wedge C_{p-2}  
\end{equation}
where $H=dB$ is the NS 3-form field strength.  The Bianchi identity $ddC_p=0$, which is equivalent to the equation of motion in Eq.~(\ref{bianchi}), implies that the gauge invariant field strengths are $(d-H)$ closed.  We will say that a configuration is topologically trivial\footnote{The K-theory classification is not sensitive to the generalized theta angles.} if the $q$-form connections $C_q$ are all globally defined, in which case these field strengths are $(d-H)$ exact.  Thus the gauge invariant field strengths $G_p$ are classified by $H$-twisted cohomology, which is the cohomology with respect to the differential operator $d-H$.

The second is the Page field strength $F_{p+1}=dC_p$, which classically is classified by de Rham cohomology as a result of the Bianchi identity $dF_{p+1}=0$ and the fact that $C_p$ is globally well-defined if $F_{p+1}$ is exact.  The well-definedness of D$p$-brane partition functions leads to a Dirac quantization condition\footnote{Although the derivative of the D-brane partition function actually contains the gauge invariant field strength $G_{p+1}$, a bulk contribution \cite{Wati} to the anomaly cancels the extra term leaving the anomaly equal to $exp(i\int(F_{p+1}))$.}, and so $F_p$'s must represent integral cohomology classes.  Eq.~(\ref{bianchi}) together with $H\wedge H=0$ implies the descent relation
\begin{equation}
H\wedge F_{p-1}=dG_{p+1}.
\end{equation}
As $G_{p+1}$ is gauge invariant, the right hand side is exact and so cohomologically trivial.  Therefore $F_{p-1}$ belongs to a special subset of cohomology that is annihilated by $H$.  Further if $F_{p+1}$ is equal to $H$ wedged with a globally defined, closed $(p-2)$-form $\omega_{p-2}$ then $F_{p+1}$ may be gauged away by the patchwise transformation
\begin{equation}
C_p\longrightarrow C_p+H\wedge \eta_{p-3}\sp d\eta_{p-3}=\omega_{p-2}
\end{equation}
where $\eta_{p-3}$ exists on each patch because $\omega_{p-2}$ is closed.  We have then learned that the Page field strengths correspond to those elements of integral cohomology in the kernel of $H\cup$ quotiented by those in its image, up to torsion corrections that are not apparent in a supergravity analysis.  

If we assume that the only possible field strengths on cycles are those that could be generated by filling in those cycles with a contractible space and including branes that source the flux, then we find a possible form for such torsion corrections from the Freed-Witten \cite{FW} anomaly on the branes.  Instead of taking the cohomology with respect to $H$, we must take it with respect to $Sq^3+H$ and then with respect to an unknown degree five operator\footnote{A term in this operator was calculated in Ref.~\cite{Uday} generalizing an example in Ref.~\cite{MMS}.  But it is not known whether there are further contributions.}.  This is the Atiyah-Hirzebruch spectral sequence (AHSS) construction of twisted K-theory.  The classification of Page field strengths by twisted K-theory explains the fact that gauge invariant field strengths are classified by $H$-twisted cohomology, they are the twisted Chern characters \cite{BCMMS,MathStev} of the corresponding twisted $K$-theory.  They are gauge invariant because they are uniquely determined by the $K$-class corresponding to the physical configuration.  The Page field strengths on the other hand are the integral cohomology classes that appear in the AHSS.  They are not gauge invariant because the higher differentials of this sequence identify distinct cohomology classes.  In other words the gauge transformations shift the Page fieldstrengths by terms that are exact under the AHSS differentials.

\subsection{Classifying Solitons}

In Ref.~\cite{WKTheory} Witten derived the K-theory classification of type IIB D-branes in the absence of nontorsion $H$ flux using Sen's construction \cite{Sen}.  Arbitrary D-brane configurations are encoded in stacks of D9's and $\overline{\textup{D9}}$'s with nontrivial gauge bundles.  Up to curvature corrections the difference between the $k$th Chern classes of the D9 and $\overline{\textup{D9}}$ gauge bundles yields the D$(9-2k)$-brane charge.  Thus brane configurations are classified by pairs of gauge bundles.  Tachyon condensation yields an equivalence of the pairs of gauge bundles under direct sums and so one concludes that D-branes are classified by the K-theory of the 10-dimensional spacetime.  Ho$\check{\textup{r}}$ava extended this construction to type IIA using non-BPS D9-branes in Ref.~\cite{PetrIIA}.

$p$-brane charges may be defined by Gauss' Law to be the integral of the corresponding $(8-p)$-form Page field strength $F_{8-p}$ over a linking $(8-p)$-cycle.  Thus if Page field strengths are classified by twisted K-theory, one may expect branes to be classified by some kind of twisted K-theory as well.  The fields are integrated over a cycle that links the cycle on which the brane is wrapped, which suggests that branes are classified by some kind of dual K-theory.  

In Ref.~\cite{MMS} MMS have argued that this dual K-theory should be the K-homology of a 9-dimensional timeslice.  Their answer differs from Witten's answer, but in fact they were answering a different question.  MMS considered spacetimes of the form $M^9\times \R$, where the $\R$ is intuitively considered to be the time direction but the signature plays no role.  They attempted to classify the conserved charges, that is the configurations on a 9-dimensional timeslice quotiented by the cobordisms that are dynamical processes relating configurations at two different times.  These dynamical processes are the MMS instantons that will be described below.  The fact that D-branes should be classified by equivalence classes of subsets and homology and not cohomology is a consequence of the construction.  One begins with an arbitrary wrapping of D-branes, which defines a homology class represented by the wrapped cycle.  Then one eliminates the anomalous wrappings and identifies those related by dynamical processes, leaving a quotient of a subset of homology.

On the other hand Witten classified 10-dimensional D-brane configurations, identifying those related by tachyon condensation.  However, the 10-dimensions include not only the spatial directions but also the time direction.  Thus while tachyon condensation is a dynamical process, the 10-dimensional configurations that are considered to be equivalent do not transform into each other over time, as the time direction has already been used in defining the configuration.  Rather these configurations are somehow equivalent in the infrared.  In our case we will see that they are reductions of the same M-theory configuration as seen by probes at different energy scales.

The difference between the 9 and 10-dimensional perspectives is not quite enough to explain the fact that Witten's approach concluded that D-branes are classified by a refinement of cohomology while MMS found a refinement of homology.  To see this, we may consider an example with only 3 relevant dimensions and no $H$ flux, and so that the K-theory is just ordinary integral cohomology and the K-homology is just integral homology.  For example, we may consider type IIB string theory on $S^3/\Z_3\times \R^7$ where $S^3/\Z_3$ is the circle bundle over the 2-sphere with Chern class equal to 3.  One-branes owe their stability to 
\begin{equation}
H_1(S^3/\Z_3)=H^2(S^3/\Z_3)=\Z_3.
\end{equation}
In the MMS picture D1-branes are classified by the cycles that they wrap, and so by $H_1(S^3/Z_3)=\Z_3$.  Meanwhile in Witten's picture one may ignore the six trivial time directions and consider a single D3 and $\overline{\textup{D3}}$ wrapped on the lens space $S^3/\Z_3$.  These each admit $U(1)$ gauge fields, which are characterized by Chern classes in $H^2(S^3/\Z_3)=\Z_3$.  The D1-charge is just the difference between these Chern classes, and so is $\Z_3$-valued.  Thus the two classifications appear to agree in this example.  It seems plausible that in general the Poincare duality that relates K-theory and K-homology implies that these two classification schemes agree.  An extension of this Poincare duality to twisted K-theory may reveal a similar agreement, although Witten's argument did not apply to the general twisted case \cite{Kapustin}.

\subsection{MMS Instantons and the Conifold} \label{mmssec}

If a D-brane wraps a $spin$ cycle then it supports a $U(1)$ gauge field with field strength $F$.  $F$ is quantized but is subject to gauge transformations, some of which are large, that mix it with the NS 2-form $B$.  The gauge invariant quantity is $B+F$.  Gauge invariance implies that in particular the integral of $B+F$ over any two-cycle $\Sigma$ must be single-valued, although unlike $F$, $B$ is not quantized and its integral is not invariant under deformations of the 2-cycle. 

As $B$ is not gauge invariant, the integral of $B$ over $\Sigma$ is not well-defined.  However if the brane's worldvolume contains a 3-manifold $M^3$ bounded by $\Sigma$ one may define the integral by Stokes theorem.  If there is a nonvanishing 3-cycle $Z^3\supset M^3$ then $\Sigma$ bounds (with opposite orientations) both $M^3$ and $Z^3-M^3$ and so we may define the integral by Stokes theorem in two different ways, corresponding to two gauge choices for $B$.  The requirement that these two definitions yield the same integral of the gauge invariant quantity $B+F$ is
\begin{equation}
\int_{Z^3}H+dF=\int_{M^3}H+dF+\int_{Z^3-M^3}H+dF=(\int_\Sigma B+F) - (\int_\Sigma B+F)=0 \label{classfw}
\end{equation}
where the minus sign came from the opposite orientations of the boundaries.  In the quantum theory both the integrals of $H$ and $dF$ are quantized, and so (\ref{classfw}) is the condition that the integral cohomology class $H+dF$ vanishes. 

Freed and Witten \cite{FW} have generalized this result to the case in which a D$(p+2)$-brane wraps a manifold that is not necessarily either $spin$ or $spin^c$.  In this case there is a $\Z_2$ torsion correction and the condition in integral cohomology becomes
\begin{equation}
{W_3+H+dF=0}
\end{equation}
where $W_3$ is the third Stiefel-Whitney of the normal bundle of $Z^3$.  While there is not a real gauge bundle away from the $spin$ case, $dF$ may still be defined as the Poincare dual to the endpoint of a D$p$-brane in the D$(p+2)$-brane, which is the generalization of a worldvolume magnetic monopole to the case in which the gauge ``bundle'' transition functions do not satisfy the triple overlap condition.  Thus if a D$(p+2)$-brane wraps a 3-cycle with nonvanishing $W_3+H$ then D$p$-branes must end on that D-brane, and the intersection of their endpoints with the 3-cycle must be Poincare dual to $W_3+H$.  These configurations generalize the baryons of Ref.~\cite{Baryons}.

MMS \cite{MMS}, reducing a more cumbersome construction by Diaconescu, Moore and Witten \cite{DMW}, realized that the baryon is a vertex that does not conserve D$p$-brane charge.  In particular imagine that there is a 3-cycle $Z^3$ such that the third Stiefel-Whitney class of its normal bundle plus the $H$ flux is some cohomology class\footnote{We will treat $k$ as a number, but in fact the contribution from $W_3$ is $\Z_2$-torsion.  However in the cases of interest this clarification will not be important.} $k$
\begin{equation}
W_3+H=k.
\end{equation}
Now we may scatter $n$ D$p$-branes and watch for D$p$-number violation.  For ease of description we will consider the case in which the cycle is a 3-sphere, but the general case is no more difficult, one need only use a Morse function on $Z^3$ to create cross-sections, as is done in Ref.~\cite{Nov}.  The 3-sphere, however, has the extra advantage that this story may be quantitatively studied in the case of D0-branes in the SU(2) WZW model as we review in Sec.~\ref{wzwsec}.

We will consider D$p$-branes that initially intersect the 3-sphere transversely.  A single D$p$-brane, far away from the others, is then locally just a point on the $S^3$.  The positions of two D$p$-branes are encoded in a two by two matrix and as they approach each other the potential terms that suppress the off-diagonal elements become small.  Once these potential terms are small enough that they do not dominate over the contribution of the $H$ flux, the two D$p$-branes blow up into a dielectric (anti-) D$(p+2)$-brane whose intersection with the 3-sphere is a surface.  For simplicity we may consider $SO(3)$-symmetric configurations and so the D$(p+2)$-brane wraps a 2-sphere at constant latitude with respect to the south pole, as such 2-spheres are invariant under the $SO(3)$ isometry.  This D$(p+2)$-brane carries no net D$(p+2)$-brane charge, as antipodal points on the $S^2$ carry opposite charges.  However its worldvolume $U(1)$ gauge bundle has a Chern class of two measured in the gauge in which the integral of $B$ over the 2-sphere is the integral of $H$ in the 3-ball to its south.  This implies that it still carries the original D$p$-brane charge.  

If instead we choose the gauge in which the integral of $B$ over the 2-sphere equals minus the integral of $H$ over the 3-disk to its north than we have performed the gauge transformation
\begin{equation}
\int_{S^2} B\longrightarrow \int_{S^2}B\p=-(\int_{S^3}H-\int_{S^2}B)=-k+\int_{S^2} B.
\end{equation}
The gauge invariance of $B+F$ then yields the transformation of the D$p$-brane charge, which is minus (because of the orientation that we have chosen for the D$(p+2)$-brane) the integral of $F$ over the 2-cycle
\begin{eqnarray}
Q_{Dp}&=&-\int_{S^2}F=-\int_{S^2}(B+F)+\int_{S^2}B\label{gaugex}\\
&\longrightarrow& -\int_{S^2}(B+F)+\int_{S^2}B\p=-\int_{S^2}(B+F)-k+\int_{S^2}{B}=Q_{Dp}-k\nonumber
\end{eqnarray}
and so we already see that D$p$-brane charge is only conserved modulo $k$.  We will see this more directly momentarily.  Note however that, as will be the case for the Klebanov-Strassler geometry, if this $H$ flux is sourced by a stack of NS5-branes then this same process will lead to a D$p$-charge on the NS5-brane worldvolume that precisely compensates for the D$p$-charge nonconservation above.

\begin{figure}[ht]
  \centering \includegraphics[width=6in]{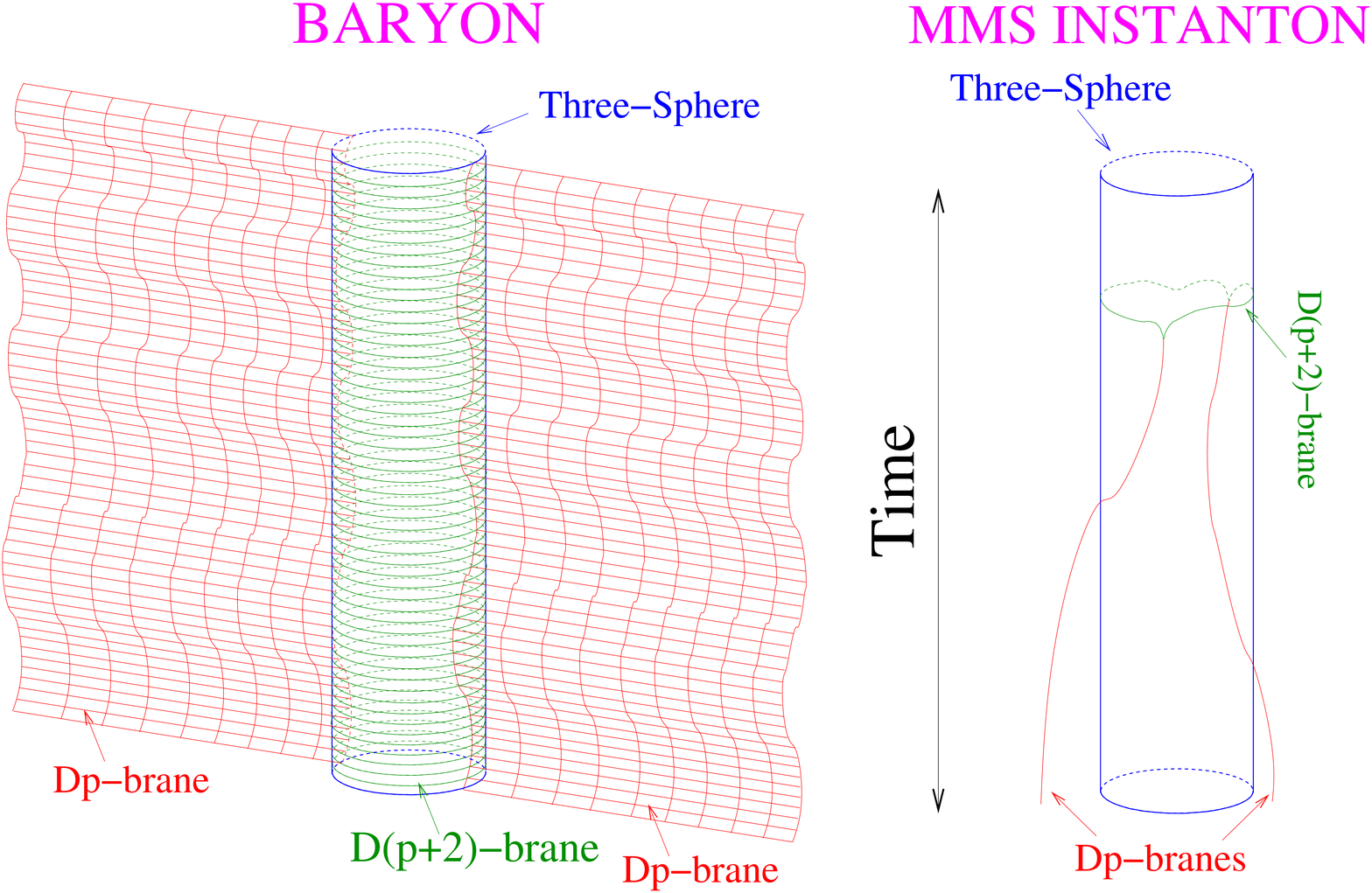}
\caption{The baryon and MMS Instanton of type II string theory on $\R^7\times S^3$ with $\int_{S^3}H=2$.  The 2 units of $H$-flux imply that 2 D$p$-branes must end on every D$(p+2)$-brane wrapping the $S^3$.  The MMS instanton violates D$p$-brane charge conservation by 2 units.  Charges are time-independent in the baryon configuration, and in fact the baryon number itself is conserved as it represents a nontrivial homology class $[1]\in \textup{H}_3(S^3)$, albeit a class that does not lift to K-homology.} \label{mms}
\end{figure}

As more D$p$-branes are added, the latitude of this two-sphere (or the value of the Morse function in the case of a general 3-cycle) increases.  For a generic number $n$ of D$p$-branes the latitude depends on specific details of the configuration.  An interesting latitude to consider is 90 degrees north.  That is to say that we add branes until our 2-cycle sweeps out the entire 3-cycle and we are again left with a stack of D$p$-branes.  Like the initial configuration, there is a natural choice of gauge for $B$, the gauge in which $B$ is finite and so its integral over a point vanishes.  However in this case that choice of gauge corresponds to integrating $H$ over the region to the north of the 2-cycle, which has no volume and so gives zero as seen in Eq.~(\ref{gaugex}).  This is not the same gauge choice in which the integral of $B$ over the south pole was zero, it differs by $k$.  Thus the integral of $F$ also differs by $k$, and $k$ units of D$p$-brane charge have disappeared.  MMS found the same answer using the baryon construction above: a D$(p+2)$-brane has swept out a 3-cycle with $k$ units of $H+W_3$ and so anomaly cancellation requires $k$ D$p$-brane insertions.  The other ends of these D$p$-branes are in the past (instead of at spatial infinity as in the baryon of Ref.~\cite{Baryons}) and so this baryon, also known as a MMS instanton, consumes the $k$ D$p$-branes.  Both configurations may be found in Fig.~\ref{mms}.

In addition to consuming D$p$-branes the MMS instanton also leaves behind a residual flux.  The D$(p+2)$-brane wrapped a noncontractible cycle and so, like an electron and positron that relatively wrap a circle and then annihilate, it leaves minus one unit of $F_{6-p}$ RR field strength in the place of the D$p$-branes that it has annihilated.  The residual Page field strength is essential in demonstrating the gauge-invariance of $G_{8-p}$.  Recall that the original D$p$-brane charge can be measured by Gauss' Law by integrating $F_{8-p}$ over a linking $(8-p)$-cycle $M^{8-p}$.  By Stoke's theorem, the D$p$-brane charge is equivalently measured by a violation of the Bianchi identity $dF_{8-p}=Q_{Dp}$.  Thus as the D$p$-branes are consumed, the integral of $dF_{8-p}$ over a transverse cycle changes, where the transverse cycle wraps the 3-cycle because the D$p$-branes were transverse to the 3-cycle.  This is consistent with the fact that Page charges are quantized but not conserved.  

On the other hand the gauge invariant field strength $G_{8-p}$ needs to be conserved.  In particular, the integral of $dG_{8-p}$ over any two surfaces bounded by $M^{8-p}$ must agree.  We will compare this integral on a surface that intersects the D$p$-branes with the integral on a surface at a later time when the D$p$-branes have already decayed. These two calculations of $\int_{M^{8-p}} G_{8-p}$ agree thanks to the new flux
\begin{equation}
\Delta(dG_{8-p})=\Delta(dF_{8-p})-\Delta(H\wedge G_{6-p})=-k-H\Delta(G_{6-p})=-k-k\cdot (-1)=0 \label{gaugeinvar}
\end{equation}
where we have used the fact that $H$ does not change during this process.  Equivalently the gauge invariance of $G_{8-p}$ implies the vanishing of the integral of $dG_{8-p}$ over any closed surface.  Eq.~(\ref{gaugeinvar}) is the evaluation of the integral over a surface that links the instanton.

Depending on the specifics of the geometry the MMS instanton may also create a nonquantized field strength along the noncompact directions.  While these cannot be determined from the topology alone, they play a role in the gauge theory as they determine the superpotential.

At the end of this process we have a stack of D$p$-branes, and so the same process may repeat again, and with each iteration the stack will lose $k$ units of D$p$-brane charge and also a unit of $F_{6-p}$ flux.  In the SU(2) WZW model this process recurs until the total brane charge is less than $k$ units.  We claim, generalizing the claim of Ref.~\cite{KPV}, that an example of this series of processes in which D$p$-brane charge is lost and $F_{6-p}$ flux is gained is S-dual to the KPV description of the Klebanov-Strassler cascade.  At each step of the cascade $k$ D3-branes in a $G_3$ background blow up into a dielectric NS5-brane which sweeps out the 3-cycle of the deformed conifold and disappears, leaving behind a unit of $H$ flux along the 3-cycle and nonquantized $H$ flux in the other directions that determines the superpotential.  The $G_3$ background is created by 5-branes wrapping a contractible trivial 2-cycle.  Because this 2-cycle is contractible the $G_3$ flux that it creates does not represent a cohomology class of the conifold, rather the $G_3$ flux felt by a D3-brane depends on the cycle it wraps and also its linking number with the D5-branes.  These linking numbers, in turn, are determined by the choice of the vacuum in the low energy effective gauge theory.  The baryonic vacua correspond to D3-branes that always link all $m$ of the D5-branes and so the rank of the gauge groups falls by $m$ at each step in the cascade.   

\subsection{More Baryons}
One may, as in Ref.~\cite{Baryons}, consider wrapping other things over cycles in the conifold to look for varieties of baryons.  For example one may wrap a D5-brane around the $S^3\times S^2$, which supports $n$ units of $G_5$ flux.  The $C_4\wedge B$ term in the D5-brane worldvolume action then implies that $n$ fundamental strings must end on the D5.  If these strings have a finite energy and so a finite length then the other ends must be on the branes in which the $SU(n)\times SU(n+m)$ gauge theory lives, and so we have found a vortex of $n$ particles in the gauge theory.  Numerology suggests that in the worldvolume gauge theory this is a kind of baryon vortex coupling $n$ external charges that each transform in the fundamental representation of $SU(n)$.  In the Klebanov-Strassler geometry the conifold is deformed, which means that a sufficiently low-energy probe notices that the 2-sphere is contractible and so this D5-brane carries no net charge.  Thus this description of the baryon breaks down at low energies, as may be expected from the fact that the $SU(n)$ gauge group does not survive the cascade.

One may also wrap a D3-brane around the 3-sphere.  The 3-sphere supports $m$ units of $G_3$ flux and so $m$ fundamental strings must end on the D3-brane.  In addition there is an energy-dependent amount of $H$ flux on the 3-sphere, which suggests that some number of D-string insertions may be required as well.  In the gauge theory there then appears to be a vortex coupling $m$ external charges, with possibly an energy-scale dependent number of magnetic monopoles.  Such a vortex may be constructed by contracting the epsilon tensors of the $SU(n)$ and the $SU(n+m)$ gauge groups, in other words it is a kind of bound state of an $SU(n+m)$ baryon and an $SU(n)$ antibaryon.  These baryons have appeared in Ref.~\cite{KS}.

So what about the $SU(n+m)$ baryon?  It must somehow be the sum of the two baryons just constructed, but on the other hand it would be surprising if it looked very different from the $SU(n)$ baryon.  To superimpose the above two baryons, one need only note that the D3-brane charge of the 5-brane is just the integral of the U(1) field strength over the 2-sphere.  The normal bundle of the 2-sphere is not $spin$, and so the integral of the field strength obeys the shifted quantization condition
\begin{equation}
\int_{S^2}F\in\Z+\frac{1}{2}.
\end{equation}
This shifted quantization condition is responsible, for example, for the quantum Hall effect that leads the fermions of Ref.~\cite{MN} to exhibit bosonic statistics.   Thus the D5-branes always carry half-integral D3-brane charge.  The $SU(n+m)$ baryon is a D5-brane whose worldvolume U(1) gauge field has a Chern class one higher than the $SU(n)$ baryon, while the $SU(n-m)$ baryon has a Chern class that is one lower.  All of these baryons exist at every energy scale, as a bundle exists with every half-integral Chern class.  In the gauge theory, these are all just boundstates of the $SU(n)$ baryon and some number of $SU(m)$ baryons.  One cannot try to calculate the flux on a D5-brane that corresponds to a specific baryon, because $F$ is subject to large gauge transformations.  $F+B$ is gauge invariant, and a natural choice for $B$ shifts by a unit each time an NS5-brane passes by, that is each step in the cascade.  And so at each step in the cascade the amount of flux corresponding to each choice of baryon shifts by one, which compensates for the shift in the gauge group.  

Notice that the mysterious D-strings ending on the $SU(m)$ baryons must end on the NS5-branes, because of the $G_5$ flux created by the D3-branes and the $C_4\wedge C_2$ coupling on the NS5-brane worldvolume.  Thus at each step in the cascade, when an NS5-brane passes through, one of these D-strings shrinks to zero length and then never re-emerges after the NS5-brane has passed.  This is just the Hanany-Witten transition \cite{HanWit}.  Then the number of monopoles coupled by the vortex at a given energy scale only depends on the difference between the rank of the vortex and the rank of the gauge group at that energy scale and not on any arbitrary initial choice of energy scale.  Up to a constant shift, there is one monopole for each $m$-index epsilon tensor contracted with the original $n+m$-index tensor.  This may be seen directly from the
\begin{equation}
S_{D5}\supset\int C_2\wedge B\wedge (B+F)
\end{equation}
term in the D5-brane worldvolume action.  The integral of one $B+F$ over the 2-sphere counts, up to a constant and typically nonintegral shift, the number of $m$-epsilon tensors contracted with the epsilon tensor of rank equal to the rank of one of the gauge groups at the energy scale at which $B$ is measured.  The $C_2\wedge B$ term then, as usual, implies that for each unit of $H$ flux there is a D-string ending on each of these D3-branes, which agrees with the fact that for each unit of D3-brane charge there is a D-string ending on each NS5-brane.  

\section{Dielectric NS5-Branes from the $SU(2)$ WZW Model} \label{wzwsec}


The $SU(2)$ WZW model at level $m-2$ describes string theory on the $SU(2)$ group manifold with $m$ units if $H$ flux.  Using boundary conformal field theory a number of authors have shown (see for example \cite{AlexSchom} for a description of the final states) that if a stack of $n$ D0-branes is placed on the 3-sphere then the $H$ flux causes the D0-branes to fatten into dielectric D2-branes that wrap a 2-sphere at a constant latitude.  The latitude is only known in the semiclassical (large $m$) regime \cite{BDS}, where it has been found to be $n\pi/m$ using the Born-Infeld action on the D2-brane.  However the boundary CFT approach holds for any value of $m$ and demonstrates that as the number of D0-branes increases, the latitude increases and upon adding the $m-1$st brane the spherical D2-brane collapses to an anti-D0-brane at the north pole.   If we add one more D0-brane, for a total of $m$, it annihilates this anti-D0 and no branes remain.  If instead we started with $m+k$ D0-branes we would then find that $m$ of them sweep out the 3-sphere and then self-destruct as above, leaving $k$, which themselves form a spherical D2-brane.  If $k>m$ then the new D2-brane sweeps out the sphere again and so on.  Thus we find a cascade, the D0-branes blow up into spherical D2-branes which sweep out the sphere repeatedly until the D0-brane charge is less than $m$, at which point a single spherical $D2$-brane remains.

The $SU(2)$ WZW model alone does not describe a critical string theory, however it may be a component of a critical string theory.  For example the near-horizon limit of a stack of $m$ NS5-branes and F-strings wrapped on a 4-torus yields string theory on $AdS^3\times S^3\times T^4$ with $m$ units of $H$ flux supported on the $S^3$.  The $S^3$ component is then described by a level $m-2$ $SU(2)$ WZW model, even away from large $m$ and without taking the near-horizon limit.  One can then watch the cascade.  For example, one may place a stack of $n$ D$p$-branes at a point on the $S^3$ with any extra directions extended along the $T^4$ and watch the system relax.  The D$p$-branes will blow up into dielectric D$(p+2)$-branes and sweep out the 3-sphere as many times as they can, each time losing $m$ units of D$p$-brane charge.  The worldvolume $SU(n)$ theory on these D$p$-branes will then cascade, although the extra D$p$-charge of the NS5-branes means that we need to be careful when calculating the effect on the gauge group.

Instead of watching the cascade occur over time, one may watch it occur as the radius varies.  That is one may consider a time-independent configuration in which the D$p$-branes extend from the horizon (or from infinity is we do not take large $m$ and the near-horizon limit).  One may then take cross-sections at different radii and compare the D$p$-brane configurations.  To do this one must specify some kind of boundary condition, we will choose a condition in which at some fixed radii the cross-section contains $n$ D$p$-branes that are on top of each other and not blown-up into dielectric D$(p+2)$-branes.  The energy of a configuration at a smaller radius is then just the energy from the WZW-model plus a correction from the derivative of the configuration with respect to the radius, which is negligible so long as the configuration relaxes slowly enough with respect to the radius.  Thus as we move in closer to the origin, we see the D$p$-branes blowing up and sweeping out the 3-sphere again and again until finally they carry less than $m$ units of D$p$-brane charge, at which point they form a spherical D$(p+2)$-brane that roughly sweeps out a cone as it dives into the origin.  From the point of view of the worldvolume theory of the NS5-brane, cross-sections of smaller radii correspond to lower energies.  Thus decreasing cross-sections correspond to flowing into the IR of the worldvolume theory, and so like the Klebanov-Strassler cascade, this cascade proceeds as one flows into the infrared.  In the large $m$\ or $n$ limit this is the familiar fact from the AdS/CFT correspondence that the IR corresponds to small radii.

Different values of $p$ are related by T-duality on the 4-torus.  If we consider $p=3$ and then S-dualize we find the desired D3-branes in a background $G_3$ flux expanding into dielectric NS5-branes and sweeping out 3-spheres.  This is not the same as the Klebanov-Strassler cascade because the spacetime contains a cone over $S^3$ and not over the 5-manifold $T^{1,1}$, however the extra 2-sphere in the conifold appears not to qualitatively interfere with the dielectric effect on the $S^3$.  In the conifold case we generally adopt the perspective in which we watch the system relax in time rather than using boundary conditions to orchestrate the evolution as a function of radius that may be interpreted as an RG-flow.  However in the T-dual MQCD perspective, when we consider a finite string coupling by including the M-theory circle or by allowing the branes to bend, we see that this RG-flow is just a radial-dependence of a supergravity solution.  Thus we expect that, by T-duality, if we include finite-coupling effects by lifting the conifold to F-theory we will see the entire cascade at once.  In particular the D3-branes lift to F-theory 5-branes\footnote{While in the IIB description the dielectric effect appears to be a dramatic transformation of D3-branes into 5-branes, in this particular F-theory description like in the M-theory description one may expect a single brane.} whose embedding depends on the torus coordinates such that going about one of the cycles of the torus (the IIA cycle) the configuration shrinks while the branes sweep out the 3-sphere.

\section{The Cascade from MQCD Brane Cartoons} \label{mqcdsec}

The Klebanov-Strassler realization of the cascade using D3 and D5-branes on the conifold is T-dual to an equivalent brane cartoon in IIA.  Such brane cartoons do not require large $N$ as one may learn about the gauge theory without using the dual bulk supergravity, semiclassical information is contained in the parameters of the brane cartoon.  Quantum mechanical features of the configuration, such as the RG flow and the theta angle, are encoded in the cartoon's M-theory lift.  

The reduction of an M-theory configuration to a IIA configuration does not necessarily exist and in general is not unique.  We will see that for configurations whose low energy effective theories describe the baryonic root of superQCD the reduction is determined by the energy scale considered.  Considering progressively lower energy scales one obtains different IIA brane cartoons that correspond to different stages in the cascade.  Each of these cartoons may then be T-dualized to obtain a corresponding configuration of branes and fluxes in the Klebanov-Strassler geometry, and so one may watch the cascade progress in IIB.  In the KPV proposal this effect is replicated by introducing anti D3-branes.

The description of the IIB cascade that we find will be a cousin of the MMS instanton.  Rather than the D3-branes expanding into dielectric NS5-branes at fixed radii, which would break the symmetries of the conifold and therefore perhaps some internal symmetries of the gauge theory, we will see a symmetric RG-flow trajectory.  The RG-flow trajectory produced by MQCD and the symmetric reduction to IIA that we will choose is T-dual to a conifold geometry in which $n$ D5-branes wrapping the trivial 2-cycle contract.  In contracting they sweep out $H$ flux and so the integral of the NS 2-form $B$ over the two-cycle monotonically shrinks.  Each time that $B$ shrinks by one unit, this is gauge equivalent to the integral of the field strength $F$ of the U(1) gauge field on each D5-brane decreasing by one.  Making this gauge transformation to keep $B$ in the fundamental domain $B\in [0,2\pi)$ we see that each D5-brane loses one unit of D3-brane charge each time that a unit of $H$ flux is swept out.  The result is that $m$ units of D3-brane charge are lost at each step.  The cascade stops when there is no more $H$ flux to link.  The baryonic root corresponds to the configuration in which each D5-brane links the same maximal amount of $H$ flux, but as is clear in the M-theory picture there are many other choices of vacuum available.

\subsection{The $N=2$ Cascade in MQCD}

Consider ``weakly coupled'' type IIA string theory on $\R^9\times S^1$ where the compact direction is $x^6\sim x^6+1$.  Extend two NS5-branes along $x^{0-5}$ both at $x^{7-9}=0$ but at distinct $x^6$ coordinates $x^6=a-1/2$ and $x^6=1/2-a$.  Let $n$ D4-branes wrap the circle and also extend along $x^{0-3}$ while another $m$ are parallel but only extend from $x^6=a-1/2$ to $x^6=1/2-a$.  For now we will place all of the D4-branes at $x^{4,5,7-9}=0$.  

Open strings may connect various D4-branes and in a limit in which the string coupling and the circle are small these decouple from other excitations and describe a four-dimensional $N=2$, $SU(n)\times SU(n+m)\times U(1)$ gauge theory \cite{WittenMQCD}.  In fact even the $U(1)$, which describes center of mass motions of the system of D4-branes, decouples and we will often omit it.  Fundamental strings connecting D4-branes without crossing NS5-branes yield the vectormultiplets.  In addition there are 2 bifundamental $SU(n)\times SU(n+m)$ hypermultiplets described by F-string-like objects (their M-theory lifts are discs and so do not wrap the entire M-theory circle) connecting D4-branes on opposite sides of an NS5.  
\begin{figure}[ht]
  \centering \includegraphics[width=6in]{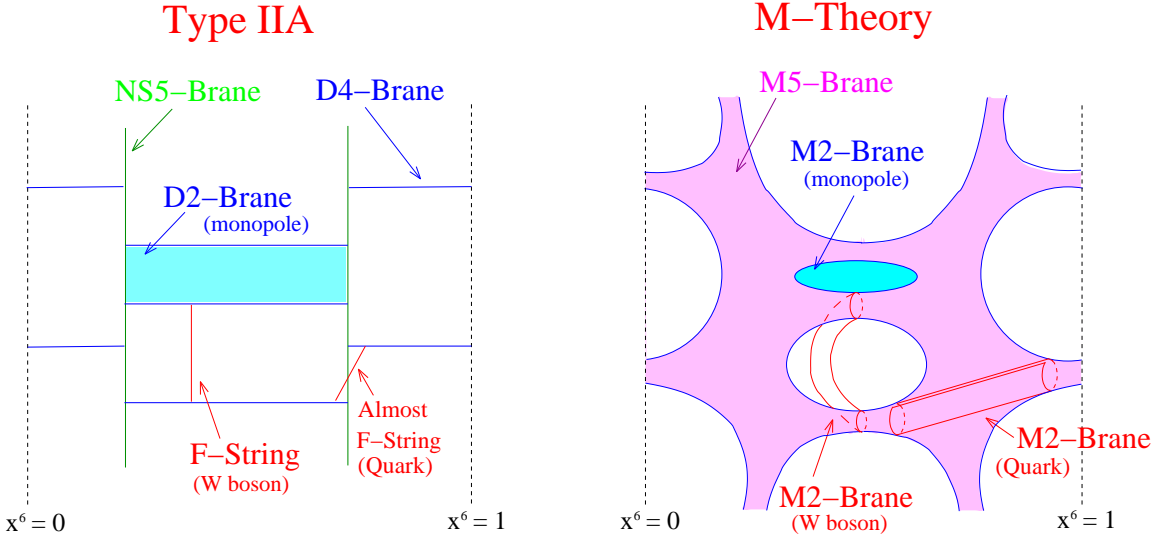}
\caption{The type IIA and M-theory realizations of a $SU(2)\times SU(3)$ $N=2$ supersymmetric gauge theory, with 2 bifundamental hypermultiplets.  In IIA the theory and vacuum are described by D4 and NS5-branes, while the degrees of freedom are F-strings and D2-branes.  In the M-theory lift the theory and vacuum are described by an M5-brane while the degrees of freedom are M2-branes.  Here all D4-branes end on NS5-branes, but in general a D4-brane may also wrap the circle $x^6$.} \label{mqcd}
\end{figure}

Quantum features of the gauge theory arise from $g_s$ effects in the brane cartoon, such as the fact that the D4-branes pull on the 
NS5-brane.  To include these effects we will lift our configuration to M-theory, and so the D4-branes and NS5-branes will be replaced by M5-branes.  This lift is not unique as two D4-brane segments separated by an NS5-brane may either connect to each other or else one connects to the NS5-brane while the other connects to another adjacent D4-brane.  These choices correspond to the choices of different vacua in the gauge theory.  Here we will consider the two simplest choices, the $r=0$ nonbaryonic root in which a minimal number of D4-branes are connected across NS5-branes and also the case with the maximal number of connections, which leads to a cascade of Seiberg dualities in which the baryonic root is selected every time.  In the $N=2$ theories considered here there are continuous moduli spaces of vacua, but these vacua are interesting as they will be among a finite collection of vacua that survive the superpotential perturbations that we will consider later. 

We will describe the M-theory lifts in terms of the complex coordinates
\begin{equation}
t=\textup{exp}(x^6+\frac{ix^{10}}{R})\sp v=x^4+ix^5
\end{equation}
where $x^{10}$ is the M-theory circle, which has radius $R$.  As the $x^6$ coordinate is periodic, different values of $t$ will correspond to the same physical point.  All of the M5-branes extend along the gauge theory directions $x^{0-3}$, and so configurations will be determined by the additional Riemann surface that they wrap in the $t-v$ plane.  

The $r=0$ nonbaryonic root vacuum consists of $n+1$ M5-branes.  $n$ of these are flat and wrap the torus $v=0$, while the last wraps the Riemann surface 
\begin{equation}
t^2+(v^m-2(-\Lambda)^m)t+\Lambda^{2m}=0\sp |t|< e^{1/2} \label{nonbar}
\end{equation}
where the bound on $t$ will be explained below.  The maximal cascade configuration is quite similar.  Again one M5-brane wraps the logarithm of the Seiberg-Witten curve
\begin{equation}
t^2+(v^m-2(-\Lambda)^m)t+\Lambda^{2m}=0\sp |t|< \textup{exp}(\frac{\lfloor\frac{n}{m}\rfloor+1}{2}) \label{bar}
\end{equation}
where $\lfloor x\rfloor$ is the greatest integer less than or equal to $x$.  In addition
\begin{equation}
k=n-\lfloor\frac{n}{m}\rfloor m
\end{equation}
M5-branes wrap the torus $v=0$.  We will argue that if a superpotential polynomial in the adjoint chiral multiplet is added then the cascade results in an $N=4$ $U(k)$ gauge theory.

The couplings of the gauge theories are inversely proportional to the square of the distance between the NS5-branes in $x^6$, and so small couplings correspond to $\Lambda<<1$.  In this limit it is possible to reverse the above process and dimensionally reduce each M-theory configuration to a IIA brane cartoon at $g_s=0$.  One may then read the gauge couplings, the ranks of the gauge groups and other semiclassical details of the theory off of the IIA brane cartoon.  The complication is that the logarithmic shape of the M5-brane means that the $x^6$ positions of the resulting NS5-branes are ill-defined.  However given a distance scale $E$, which corresponds to an energy scale proportional to $E$ because the probe M2-branes have a fixed tension, one may define a reduction to IIA as follows \cite{Hitoshi}.  Consider the points on the M5-brane that are at $|v|=E$, these consist roughly of two circles in the above limit $\Lambda<<1$, where we ignore the gauge theory directions.  These circles are at roughly fixed values of $x^6$.  We place the two NS5-branes at these two values of $x^6$ and consider the M5-brane at smaller values of $v$ to be D4-branes connecting those two NS5-branes at $v=0$.  The reduction of the flat toroidal components of M5-branes is well defined, they yield D4-branes wrapping the entire $x^6$ circle. 

\begin{figure}[ht]
  \centering \includegraphics[width=5in]{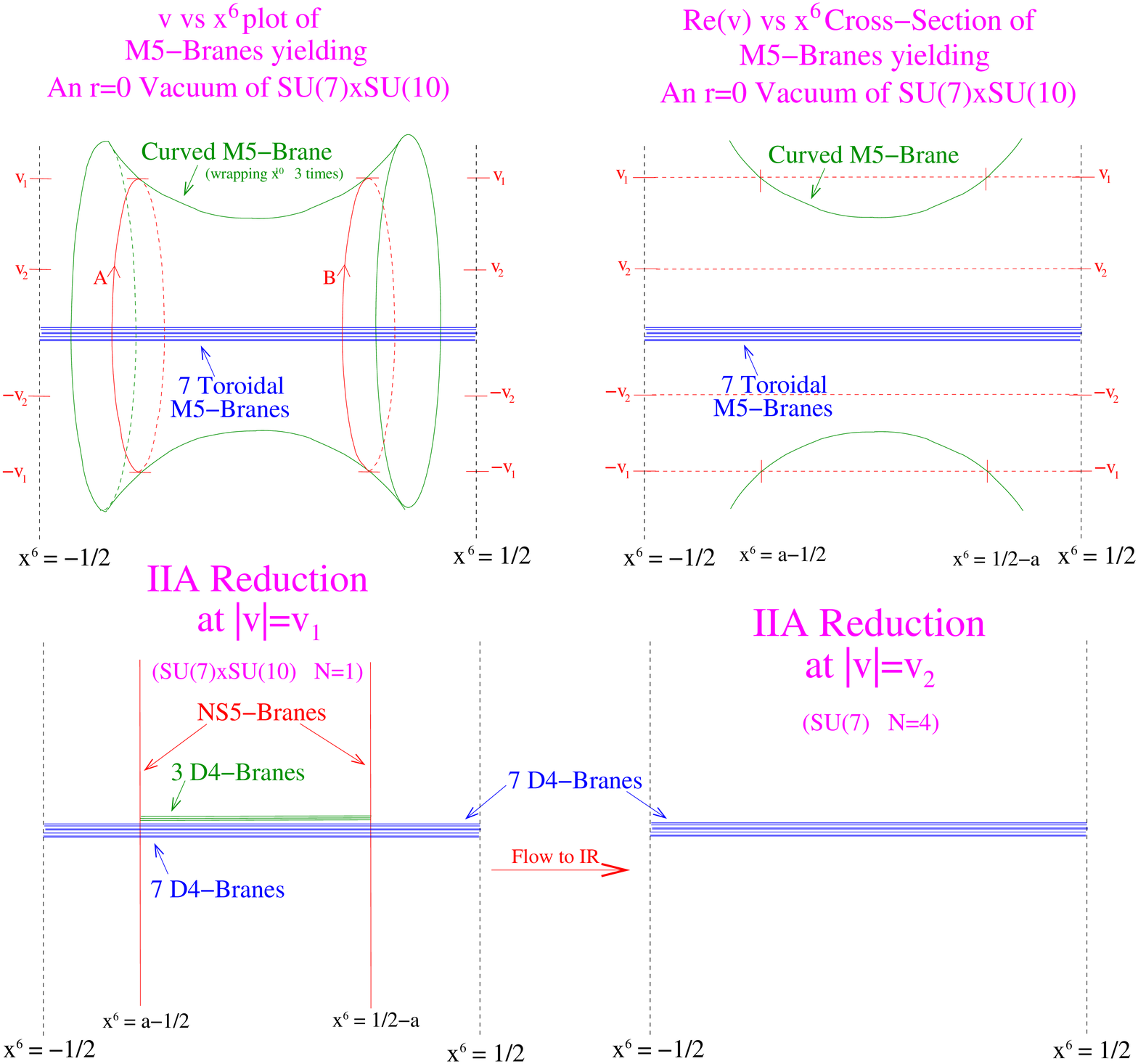}
\caption{The type IIA and M-theory realizations of the $SU(7)\times SU(10)\longrightarrow SU(7)$ cascade.  This is not a Klebanov-Strassler type cascade, which would end with only the decoupled $U(1)$ as seen in Fig.~\ref{ks}.  On the upper-left $v$ and $x^6$ are represented, while the upper-right is a projection of the same configuration onto Re($v$) and $x^6$.  The reduction to IIA is performed at two energy scales, $v_1$ and $v_2$.  The intersection of the M5 with $|v|=v_1$ consists of the loops $A$ and $B$, which have winding numbers of 3 and $-3$ about the M-theory circle.  Thus the reduction at that energy scale yields 2 NS5-branes, one connected to 3 D4's on the right and the other to 3 D4's on the left.  The M5-brane does not intersect $|v|=v_2$, and so a reduction at that energy scale does not result in any NS5-branes.  In every cartoon we see the 7 flat branes that support the $SU(7)$ $N=4$ theory at the bottom of the cascade.} \label{r0}
\end{figure}

As seen in Ref.~\cite{WittenMQCD} the distance between the NS5-branes, which determines the gauge coupling, depends logarithmically on the energy considered and so the RG flow is reproduced.  However when these two circles pass through each other the number of resulting D4-branes changes so that Seiberg duality is also reproduced \cite{Hitoshi,SeibergDuality}.  The bounds on $t$ in Eqs.~(\ref{nonbar},\ref{bar}) can now be seen to reflect the fact that given the gauge theory at some energy scale one can only determine (up to corrections that are exponentially suppressed in the allowed region) the part of the curve at lower values of $t$, which correspond to flowing the theory into the IR.  An attempt to continue into the UV is obstructed by the fact that, as one of the two gauge groups is strongly coupled in the UV, one needs to determine a UV completion.  This UV completion is not unique, but rather depends on the connections of D4-branes to NS5-branes at higher energy scales than the one in which we specified our initial conditions.  That is to say that two different vacua in two different UV theories may flow to our starting point.

In general any UV completion involves an infinite number of cascades.  An infinite number of cascades would be avoided if, for example, $m$ D4-branes connect to the NS5-branes far away on the $v$-plane, corresponding to $m$ extra vector multiplets with large adjoint scalar VEVs that in the UV combine with our $SU(n)\times SU(n+m)$ to yield the conformal theory
\begin{equation}
SU(n)\times SU(n+m)\times U(1)^m\stackrel{UV}{\longrightarrow}SU(n+m)\times SU(n+m).
\end{equation}
However such a completion could be apparent in the IR, as the $m$ extra vector multiplets would be massless although weakly coupled to the rest of the theory.  If some of the UV theories are in nonbaryonic root vacua then the number of extra D4-branes required to stop the cascade will generically be different from $m$.

We may now apply the above dimensional reduction procedure to understand the RG flows of the vacua (\ref{nonbar}) and (\ref{bar}).  In the first case, which corresponds to the $r=0$ nonbaryonic root of the $SU(n+m)$ $N=2$ gauge theory with $2n$ flavors of hypermultiplets, the circles corresponding to any $v$ in the allowed domain never cross, and so at every energy down to the strong coupling scale $E$ (defined to be the shortest distance between the curved and flat M5-branes) we find two NS5-branes connected by $m$ D4-branes and another $n$ D4-branes that wrap the entire circle.  In IIA we may equivalently say that $n$ D4-branes run between the two NS5-branes in one direction while $n+m$ run in the other direction, and so we have the original $SU(n)\times SU(n+m)$ theory.  At the energy scale $E$ the $SU(n+m)$ is strongly coupled as the corresponding NS5-branes coincide.  

It is tempting to then claim that at smaller energies, smaller than the minimal radii of the circles, there are no NS5-branes and so we are left with the $SU(n)$ theory on the $n$ D4-branes wrapping the circle.  However the circles cease to be approximately circular and in fact extend to arbitrarily small $v$ ($E=0$ in the $N=2$ case) at these energies and so this argument fails.  The curve becomes singular as the cycles that bound massless dyons degenerate.  Yet if, as seen in Fig.~\ref{r0}, we turn on a superpotential for the adjoint chiral multiplet, which gives VEVs to these massless dyons\footnote{At the baryonic root VEVs are given to a maximal set of these massless dyons, deforming away all of the singularities.}, then the circles no longer extend to arbitrarily small $v$ and one may conclude that the $SU(n+m)$ gauge symmetry, after it becomes strongly coupled, is gapped (is Higgsed by the dyon VEVs)
\begin{equation}
SU(n)\times SU(n+m)\stackrel{r=0}{\longrightarrow} SU(n).
\end{equation}
This is as one expects for the $r=0$ vacuum, after all $r$ is the rank of the remaining gauge group in the $SU(n+m)$.  This $SU(n)$ gauge theory lives on the worldvolume of the D4-branes that wrap the torus, and the light degrees of freedom do not couple to the far away 5-branes.  Thus the supersymmetry is enhanced to that of a stack of flat, parallel branes, $N=4$ \cite{KS}.  In fact the final gauge symmetry is $U(n)$, we have been consistently ignoring the extra $U(1)$ that describes the center of mass of the D4-branes.

\subsection{The $N=1$ Cascade}

We will now, as described above, introduce a nondegenerate superpotential $W(\Phi)$ for the adjoint chiral multiplet $\Phi$, thus giving VEVs to the massless dyons, resolving the singularities and allowing reductions to type IIA.  By nondegeneracy we mean not only that the extrema are distinct, but that the distances between the extrema are larger than the energy scale considered.  These two conditions ensure that it suffices to expand the superpotential to second order, and so the physics is identical to that of the $N=2$ theory with supersymmetry softly broken by a (possibly infinite) mass term for the adjoint chiral multiplet.  

\begin{figure}[ht]
  \centering \includegraphics[width=5in]{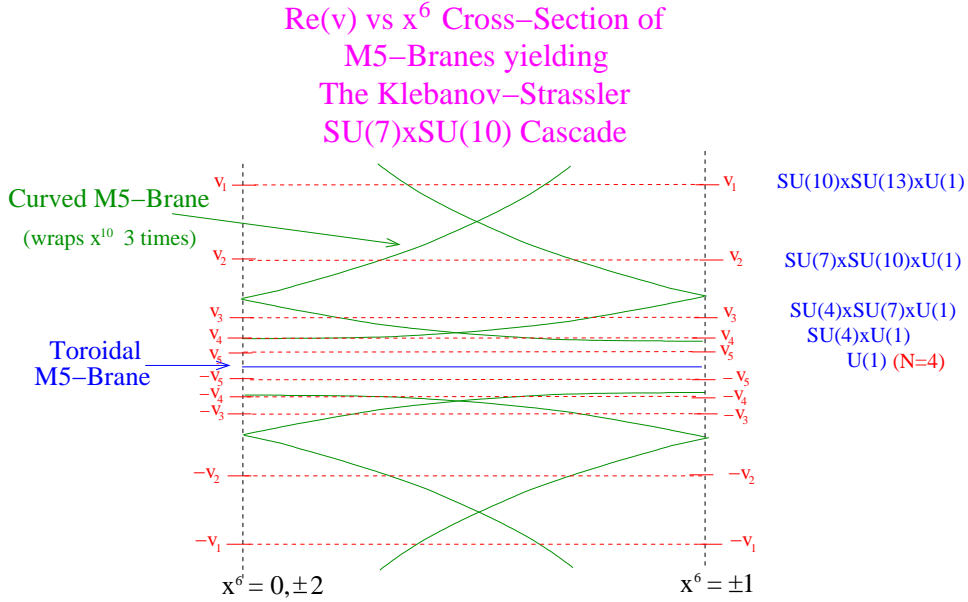}
\caption{Here we see the M-theory configuration corresponding to the Klebanov-Strassler cascade $SU(7)\times SU(10)\longrightarrow U(1)$, where the $U(1)$ is present at every stage but generally omitted.  The continuation to the higher energy scale $v_1$ is not unique, and we have chosen the continuation such that the UV $SU(10)\times SU(13)$ is in a baryonic root vacuum.  The effective gauge group at each energy scale  $v_k$ is written next to the energy scale.  When the group is $SU(a)\times SU(a+b)$ then the IIA reduction consists of two NS5-branes with $a$ D4-branes connecting them on one side and $a+b$ on the other.  Notice that this M5-brane configuration is different (only 1 flat brane) from the M5-brane configuration in Fig.~\ref{r0}, which describes a different vacuum and so a different cascade of the same initial $SU(7)\times SU(10)$ theory.} \label{ks}
\end{figure}

Introducing the new complex coordinate $w=x^8+ix^9$ the effect of a superpotential is merely to change the $x^8$ and $x^9$ coordinates of the NS5-branes from $w=0$ for both to
\begin{equation}
w=0\hsp{.3} \textup{and}\hsp{.3} w(v)=W\p(v)
\end{equation}
respectively.  In particular a superpotential does not affect, in our small $\Lambda$ approximation, the pattern of $t$ coordinate crossings.  The superpotential does have the effect of lifting most of the moduli space, leaving adjoint scalar VEVs either at the extrema of the superpotential, at which $w(v)=0$, or else at the values of the bare masses of the flavors to which they are locked.  In the first case the analysis proceeds as above but now with a potential cascade at each critical point.  The compactness of $x^6$ means that the second case corresponds to flat D4-branes wrapping the entire configuration but not ending on an NS5-brane.  While such branes contribute to the rank of the gauge group, they do not affect the cascade as they are not connected to the participating branes.  The $N=1$ case has the advantage that all of the vacua except for the $r=0$ vacua have some form of multi-step cascade, unlike the $N=2$ case above in which cascades only occurred at very special points in a huge moduli space.

Klebanov and Strassler \cite{KS} have treated the case in which the NS5-branes are orthogonal.  This is a limiting case of the above construction in which a mass term is included for the adjoint chiral multiplet $W(\Phi)=\mu\Phi^2$ and then the mass $\mu$ is taken to infinity.  

Now we are ready to study the curve Eq.~(\ref{bar}), drawn in Fig.~\ref{ks}, which has been seen in Ref.~\cite{KS} to yield the Klebanov-Strassler cascade.  This claim is easily reproduced by the above algorithm.  For concreteness, we will describe a cascade that begins with a $SU(7)\times SU(10)$ gauge theory, so that
\begin{equation}
n=7\sp m=3\sp k=1\sp |t| < e^{3/2}.
\end{equation}
Further we consider a small adjoint scalar mass $\mu$ so that the projection onto the surface $w=0$ is just (\ref{bar}).  This choice does not affect any qualitative features of what follows, and in fact the opposite limit has already been used in Ref.~\cite{KS}.  At every energy scale there will be $k=1$ D4-brane wrapping the $x^6$ circle, corresponding to the flat M5-brane component not connected to the Seiberg-Witten curve.  Following the above algorithm, we find the positions of the other branes by looking for circles at fixed $|v|$ on the M5-brane component
\begin{equation}
t^2+(v^3+2\Lambda^3)t+\Lambda^{6}=0
\end{equation} 
which may be solved for $t$
\begin{equation}
t=-\Lambda^3-\frac{v^3}{2}\pm\frac{\sqrt{(v^6+4v^3\Lambda^3)}}{2}.
\end{equation}
The M5-brane curve enjoys a $x^6\leftrightarrow -x^6$ symmetry which translates to $t\leftrightarrow 1/t$.  The small $\Lambda$ approximation leads one gauge group to be weakly coupled while the other undergoes Seiberg duality and may be used to eliminate $\Lambda$ at $|t|>1$.  Thus we may find the NS5-brane at $|t|>1$ and use the $t\leftrightarrow 1/t$ invariance of our configuration to find the other NS5-brane.  

To zeroth order in $\Lambda$ this yields $t=-v^3$ at $|t|>1$, in accord with the fact that each NS5 is connected to 3 D4's.  Thus each absolute value of $v$, corresponding to the energy scale $E=|v|$, intersects the M5-brane on the circle $|t|=E^3$.  The NS5-branes are therefore placed at
\begin{equation}
x^6=\pm 3\textup{ln}(E).
\end{equation}
Note that the distance between the two NS5-branes reproduces the known RG flow of the $N=2$\ $SU(n+m)$ theory
\begin{equation}
\frac{1}{g_{YM}^2}\sim \Delta x^6=3\textup{ln}(E)-(-3\textup{ln}(E))=6\textup{ln}(E)=(2(n+m)-2n)\textup{ln}(E)=(2N_c-N_f)\textup{ln}(E)
\end{equation}
as expected from the fact that at scales greater than $\mu$ this theory is very close to the $N=2$ theory.  The NS5-brane at positive $x^6$ is connected to $m=3$ D4-branes, all on its left, while that at negative $x^6$ is connected to 3 D4-branes on its right.  These are the same 3 D4-branes, but the number of times that the D4-branes wrap the $x^6$-circle before connecting depends on $|v|$ because the total number of D4-branes at any cross-section of constant $x^6$ must be $m=3$ times the number of M5 windings with radius less than $|v|$ plus the number of flat M5's.  In the original theory, which corresponds to $1<|x^6|=3ln(E)<3/2$ or equivalently $e<|t|<e^{3/2}$ (and its reciprocal) we find that the 3 D4-branes are emitted from one NS5-brane, make two full circles around $x^6$ and then finally continue a little farther to end on the other NS5-brane.  Thus in addition to the omnipresent disconnected D4-brane, there are an extra 9 D4-branes in the region that the 3 D4's wrap three times and an extra 6 in the region wrapped only twice.  We then find that one region contains $7=1+6$ D4-branes and the other $10=1+9$, reproducing the original $SU(7)\times SU(10)$ symmetry.  

At $|t|=e$ the NS5-branes cross and so at $\sqrt{e}<|t|<e$ the 3 D4-branes only wrap the entire $x^6$ once.  Thus the region in which there were always 7 D4-branes has shrunk, but still exists leaving an $SU(7)$.  However the region wrapped 3 times, which was between the two NS5-branes that have now passed through each other, has now ceased to exist.  Once the NS5-branes have passed each other a new region opens up between them.  This one has again the omnipresent D4, but only 3 other D4's, which are the stack of D4's that wrap the entire circle.  The remaining gauge symmetry is now $SU(4)\times SU(7)$. 

At $|t|=\sqrt{e}$ the NS5-branes cross again and so at $|t|<\sqrt{e}$ the 3 D4-branes do not wrap the entire $x^6$ circle, but merely extend directly from one NS5-brane to the next as in the $r=0$ configuration.  Thus there are two regions, one with 4 D4's through which the stack of 3 passes, and the other with only the free D4-brane, yielding a $SU(1)\times SU(4)$ symmetry.  $SU(1)$ is just a point, however in fact we have been consistently ignoring the $U(1)$ whose scalar VEV parametrizes center of mass motions of the free D4-brane, and so the symmetry is $U(1)$.  Now we have returned to the $r=0$ vacuum case.  At energy scales yet lower, so that no probe is energetic enough to extend from the toroidal M5 to the curved M5's, the $SU(4)$ gauge symmetry is Higgsed by dyon VEVs and only the $U(1)$ $N=4$ theory remains.  

\subsection{Generalization:  Gauge Groups with More Simple Components}

Adding extra NS5-branes leads to cascades with more than two gauge groups, but each iteration in the cascade still corresponds to two NS5-branes crossing and so the remaining gauge groups may be found as above.  In particular when each gauge group becomes strongly coupled then at the baryonic root the new gauge group is just the Seiberg dual, while nonbaryonic roots correspond to some D4-branes being suspended between the coincident NS5-branes in the M-theory lift and so the corresponding colors do not appear in the lower energy effective theory.  The term ``Seiberg dual'' is used loosely here, as in the case of noninfinite chiral adjoint multiplet mass (which is unavoidable in the case of an odd number of NS5-branes) the superpotential contains quartic terms that were not in Seiberg's original magnetic theory (see for example Ref.~\cite{Ken}).

The simplest class of examples of cascades with three gauge groups occurs when two NS5-branes are connected to the same number $m$ of D4-branes on the same side.  In particular if baryonic roots are chosen in every vacuum then two NS5 branes remain parallel while the third crosses them both in the same order repeatedly until the end.  Then at each step of the cascade the rank of the largest gauge group is reduced by $3m$
\begin{eqnarray}
SU(n)\times SU(n+m)\times SU(n+2m)&\longrightarrow& SU(n)\times SU(n+m)\times SU(n-m)\\&\longrightarrow& SU(n)\times SU(n-2m)\times SU(n-m)\longrightarrow ...\ .\nonumber
\end{eqnarray}
The case $m=1$ would at fairly low energies give an $N=1$ theory with a $SU(3)\times SU(2)\times U(1)$ gauge symmetry, although with the wrong matter content.  

More generally the period of the cascade is roughly the least common multiple of the ranks of the simple gauge groups times the number of simple groups.  For example if we begin with $SU(n)\times SU(n+1)\times SU(n+3)$ with the $SU(n)$ group just below its UV fixed point and the $SU(n+1)$ group at its UV fixed point then the cascade consists of drops in rank of 1, 4 or 5 at a time, repeating the sequence: $5, 4, 5, 4, 5, 4, 5, 4, 5, 1$.  In this case there are three NS5-branes, one attached to a single D4-branes extending to the right ($+x^6$), one with two D4-branes extending to the right and one with 3 D4-branes extending to the left.  These numbers $1, 2$\ and $3$ are uniquely determined by the gauge group, as they are the differences in ranks between the three simple components.  As we flow to the IR the first two NS5-branes move to the right, the second with twice the velocity of the first, while the third moves left.  The rank of a gauge group drops by $4=3+1$ when the left-moving NS5, which is attached to 3 D4's crosses the right-moving NS5 that is attached to a single brane.  Similarly a rank drops by $5=3+2$ when the left-moving NS5 crosses the right-mover attached to 2 D4-branes.  A rank drops by a single unit when the right-mover attached to 2 D4-branes overtakes the other right-mover.

In the case of more than 3 simple components the number of D4-branes attached to a given NS5-brane is not uniquely determined by the gauge group, but is determined by the gauge group together with the bifundamental matter content.  The number of D4-branes attached to a given NS5 is then the difference between the ranks of two simple components such that there is bifundamental matter charged under the two components.  At each step of the cascade two NS5-branes cross, which causes the rank of a simple component of the gauge group to drop by the difference in the number of D4-branes ending on those two NS5-branes.  Here D4-branes attached to opposite sides of NS5-branes are counted with opposite signs.  We count only D4-branes whose corresponding adjoint scalar VEVs are at the same extremum of the superpotential.  Sufficiently separated minima have cascades which may be treated separately.  The treatment of nearby minima and in particular degenerate critical points of the superpotential is beyond the scope of this note.

At large $N$ the endpoints of such transitions have been found explicitly in Ref.~\cite{OT}, who also used both the MQCD approach and its T-dual IIB realization.  

\section{T-Dualizing the MQCD Cascade} \label{dualsec}


\subsection{The Cascade from a Decreasing $B$-Field}

We will now T-dualize the $x^6$ circle of the cascading IIA brane cartoons to find the cascade in IIB.  On the IIA side there are two interpretations of the cascade.  First, different steps in the cascade are different approximations of the same M-theory configuration.  Later steps in the cascade correspond to the M-theory configuration as seen by less energetic M2-brane probes.  The second interpretation is that early steps in the cascade correspond to initial conditions in type IIA string theory in a bizarre kind of weak-coupling limit where $g_s\rightarrow 0$ prevents the NS5-branes from bending, but at the same time the tension of the D4-branes causes them to contract over time, pulling the NS5's.  As we time-evolve the system the NS5-branes are pulled together, finally annihilating to leave only the D4-branes that wrap all of $x^6$.  

The M-theory configuration is contained in the F-theory configuration, one needs only to project away the IIB circle, and so the first interpretation of the cascade may allow one to find that various steps of the cascade in the Klebanov-Strassler geometry are just reductions of F-theory to IIB at different energy scales.  In fact, it has already been seen that the conifold transition is smooth when the M-theory circle is fibered over it \cite{AtiyahWitten}.  However the IIB configurations themselves, at small $N$, contain the same information (\textit{i.e.} no RG flow information) as the type IIA configurations.  That is, each frame is a semiclassical approximation of the quantum theory, where one approximates the effective theory at some energy scale.  Thus as we watch the cascade in type IIB we will pass through successive IIB approximations at $g_s=0$ each of which is obtained from the previous by relaxing the configuration.  As in the case of M-theory reductions to type IIA, such approximations do not exist in general, for example the Argyres-Douglas \cite{AD} vacua admit no such reduction and correspondingly there is no decoupling of the full six-dimensional physics of the M5-brane worldvolume that allows the kind of four-dimensional description admitted by the baryonic root vacua.

The IIA brane cartoons are independent of the first four coordinates, in which the gauge theory lives.  Therefore we need to T-dualize the remaining 6-dimensional configuration.  As we are only interested in the ranks of the remaining gauge groups and the implications for the S-dual of the twisted K-theory classification, and not in a quantitative analysis of the condensates, we will only identify the topology of the dual configuration.  

We will T-dualize our configuration by dividing it into four-dimensional slices $\R^3\times S^1$ and then dualizing each slice using the following three facts.  First, each NS5-brane that is not parallel to the $S^1$ is T-dual to a KK-monopole with respect to the T-dual circle, and so in particular $k$ NS5-branes are T-dual to a $A_{k-1}$ singularity.  If the NS5-branes are at different points in the $\R^3$ then the singularity is blown up by topological 2-spheres that semiclassically are long cylinders with the ends pinched and so the area of a 2-sphere is proportional to the length of a tube, which is the distance between the NS5-branes in the $\R^3$.  The distance between the NS5-branes along the $x^6$ circle yields a $B$-field on the blown-up sphere, even when the area of the sphere vanishes because the NS5's are coincident along the other three directions.  Second, each D4-brane that completely wraps the circle and does not extend in any of the other three directions is T-dual to a D3-brane at a point in the T-dual four-dimensions.  Finally a D4-brane extended along the circle that ends on two NS5-branes is T-dual to a D5-brane that wraps the corresponding blown-up two-sphere\footnote{If on the way from one NS5-brane to the other a D4 wraps the circle $k$ times then there are $k$ D4's that wrap the circle, and so the T-dual carries $k$ units of D3-charge.  This is why the D3-charge (and hence the rank of the gauge group) falls as the D4 unwinds.}.  The normal bundle of this two-sphere has a nontrivial second Stiefel-Whitney class which leads to a shifted quantization condition for the worldvolume $U(1)$ gauge field, this is shown to be responsible for the quantum Hall effect in the worldvolume theories in Ref.~\cite{MN}.  The shifted quantization condition implies that the D5-branes carry half-integral D3-brane charge
\begin{equation}
Q_{D3}=\int_{S^2}F\in\Z+\frac{1}{2}. \label{squant}
\end{equation}

A simple choice of 4-dimensional slices consists of one slice for each point $v$.  Thus in the case of a gauge group $SU(n)\times SU(n+m)$ each slice consist of two NS5-branes separated by $W'(v)$ in the $\R^3$ and along the circle by the inverse coupling squared of one of the simple components.  The T-dual of that slice then consists of a resolved $A_1$ singularity where the blown-up circle has an area that is roughly $W'(v)$ and a B-field proportional to an inverse coupling squared.  

In the case of the conifold such a slicing is not convenient as one of the NS5's is parallel to the slice, and so instead we consider slices $v+w=c$ where each slice corresponds to a complex number $c$.  In this case each slice is dual to an $A_1$ singularity resolved by a 2-sphere with area $|c|$.

In addition there may be D4-branes wrapping the circle at that $v$ which become D3-branes.  If $v$ extremizes the superpotential then there may also have been D4-branes stretching between the two NS5-branes, which become D5-branes wrapping the $S^2$ with a half-unit of $U(1)$ gauge flux.  There are two ways to stretch between the two NS5's, corresponding to the colors of the two simple components of the gauge group.  In IIB these correspond to the two orientations of the D5 wrapping about the 2-sphere.  

As a consistency check, we note that two D4-segments connecting the NS5-branes on both sides can combine to create a single D4-brane that wraps the circle.  T-duality must commute with this process, and indeed it does.  The two resulting D5-branes have opposite orientations and so annihilate.  However the worldvolume $U(1)$ fluxes of $1/2$ and $-1/2$ mean that the unstable $D5-\overline{D5}$ pair carries a single unit of D3-brane charge.  And so in the end a D3-brane remains, which is the T-dual of the D4-brane wrapping $x^6$ as required.

Now it is easy to watch the cascade.  As we probe lower energies the NS5-branes sweep out $x^6$, crossing each other.  In the T-dual IIB this means that each step in the cascade corresponds to a change of the $B$-field on one 2-cycle by a single unit.  At a slice corresponding to an extremum of the superpotential, such as $c=0$ in the conifold configuration, there may be $m$ D4-branes extending between the NS5's.  As the superpotential is extremized the integral of the Kahler form over the 2-cycle vanishes.  However the B-field and in particular the complex Kahler form $K+iB$ does not generally vanish and so we will say that the T-dual configuration is a D5-brane wrapped on this 2-cycle, although in light of the vanishing real area of the two-cycle one may consider this to be a fractional D3-brane.  During each step of the cascade the NS5-branes separate and come together again on the other side of $x^6$, and so the distance increases from $0$ to $1$, which is identified with $0$ because the $x^6$ direction is compactified.  Correspondingly the pullback of the $B$-field to the worldvolume of the D5-brane decreases from $0$ to $-1$, which may be identified with $0$ via a large gauge transformation if at the same time we decrease the integral of the worldvolume gauge field $F$ of each D5 by one unit.  Note that such a shift is consistent with the shifted quantization condition (\ref{squant}).  The D3-charge carried by each D5-brane is the integral of the $F$ flux over the transverse 2-cycle and so the D3-charge decreases by one unit per D5-brane, that is the total D3-brane charge decreases by $m$ units in keeping with the fact that there are $m$ less D4's wrapping $x^6$.  This yields a decrease in the rank of each simple gauge group by $m$ units and so we have reproduced the cascade.

As noted above, instead of a time-dependent process this may be treated as a radius-dependent process.  The integral of the $B$-field over the 2-sphere scales as the logarithm of the radius \cite{KN}.  As we flow into the IR the 2-sphere wrapped by the D5-brane shrinks, at large $N$ because the IR corresponds to smaller radii and at small $N$ because in the T-dual MQCD the $|v|$ coordinate of the corresponding M5-brane shrinks, and so the $B$-field felt by its worldvolume shrinks as well.

In the radial picture the nonbaryonic vacua are easier to understand.  Nonbaryonic vacua are T-dual to configurations in which $k$ D4-branes end on two converging NS5-branes in the M-theory lift, like the $m$ D4-branes at the $r=0$ vacuum studied above.  Thus when the NS5-branes converge the D4-branes are annihilated.  In IIB this means that the corresponding $k$ D5-branes do not fall to smaller radii like their compatriots that are connected across the NS5s.  Instead these $k$ D5-branes decouple at lower radii than the transition (lower than the corresponding condensate).

So what are the allowed values of $k$?  To reproduce all of the vacua we need to T-dualize each segment of D4-brane separately, so that we find $m+n$ D5-branes and $n$ anti D5-branes, each with half a unit of worldvolume field strength to reproduce the correct D3-brane charge
\begin{equation}
Q_{D3}=n\int_{S^2}(-F)+(n+m)\int_{S^2}F=\frac{-n}{2}+\frac{m+n}{2}=\frac{m}{2}.
\end{equation}
Only the $n+m$ D5-branes are dual to the D4-branes that are shrinking to zero length, and so $0\leq k\leq m+n$.  Once the probes have become so small that the $k$ D5-branes have been decoupled, $m+n-k$ D5-branes remain.  If $k>m$ then $m+n-k<n$ and so there is more tension on the NS5-branes from the outside than the remaining tension on the inside, thus they change direction and the remaining $SU(m+n-k)\times SU(n)$ theory continues to cascade.  This is the IR effective theory of the $r=m+n-k$ vacuum of the original $SU(m+n)$ with $2n$ flavors.  The allowed range of $k$ is $m\leq k\leq m+n$ and so $0\leq r\leq n$.  

If $k=m$ then $m+n-k=n$ and so once $k$ D5-branes have decoupled we are left in a conformal theory which is a possibly strongly coupled $SU(n)$ gauge theory with $2n$ flavors.  This is just the $r=m+n-k=n=N_f/2$ vacuum of the original $SU(m+n)$ theory, which is always a conformal field theory.  Finally if $k<m$ then $m+n-k>n$ and so the NS5-branes cross as in the baryonic root vacua of a $SU(m+n-k)\times SU(n)$ theory.  This leaves the Seiberg-dual $SU(n-m+k)\times SU(n)$, and so this case corresponds to the $r=n-m+k$ vacuum of the original theory.  In all three cases we have found that the $SU(n+m)$ theory flows to an $SU(r)$ theory with $0\leq r\leq n=N_f/2$ and $N_f$ flavors, where the first and third cases contained some redundant descriptions of the same vacua. We have then reproduced the known set of vacua of an $N=2$ supersymmetric $SU(n+m)$ gauge theory with $2n$\ $(2n<n+m)$ flavors when it is softly broken to $N=1$ by an adjoint scalar mass.

\subsection{The Cascade from Dielectric NS5-Branes}

How is this description of the cascade related to the KPV dielectric NS5-brane description of Ref.~\cite{KPV}?  In the current description the D5-brane wraps a cycle whose $B$-field changes by one unit.  By Stokes' theorem we may equivalently say that the D5 sweeps out a single unit of $H$ flux
\begin{equation}
\Delta\int_{S^2} B=\int_{S^2\times I_{time}} H
\end{equation}
destroying D3-charge because the D5 worldvolume coupling
\begin{equation}
S_{D5}\supset \int B\cup C_4
\end{equation}
implies that worldvolume $B$ flux carries D3 charge.  The origin of the $H$ flux in M-theory was clear, it was dual to the bending of the M5-brane.  However when we reduced to 10-dimensions it was obscured, as the bending was no longer visible in a single frame.

To see the origin of the $H$ flux, as in M-theory, we must try to understand how the configuration relaxes.  First we note that the tension of the D5-brane includes a contribution from its D3-brane charge, and so in particular is monotonically increasing in the $B$-field for positive $B+F$ and is not periodic in $B$.  This is clear, for example, from the Dirac-Born-Infeld action on the D5-brane worldvolume.  Thus any process that decreases the $B$-field on the D5-brane will reduce the energy.  In the T-dual IIA description this matches our notion of relaxing, as decreasing the B-field corresponds to contracting the D4-branes, a process which is driven by the 4-brane tension.  Thus we are searching for a process in IIB that creates $H$ flux with 2-coordinates along the wrapped 2-cycle and also one coordinate along the time direction (or along the radial description in AdS/CFT-like descriptions of the RG flow as a flow to smaller radii).  

Such a process is a Brown-Teitelboim \cite{BT} dielectric NS5-brane.  This NS5-brane nucleates as a small, spherical NS5-brane that carries no net NS5-brane charge.  In IIB NS5-branes carry $U(1)$ gauge fields and, as this NS5-brane was created from uncharged heat, the integral of this $U(1)$ gauge field over the 2-sphere must vanish.  If the NS5-brane expands and then contracts out of existence such that its worldvolume links the stack of $m$ D5-branes then the integral of $G_3$ over its worldvolume is $m$, where smaller cascades may result from NS5-branes that only wrap some of the $m$ units of D5-brane charge.  To find the $r=0$ case it suffices to probe smaller distance scales than the radius of D5's before any NS5-branes have a chance to nucleate.  Any NS5-branes nucleating at smaller scales will not link any D5-branes and so do not affect the gauge symmetry.

We will consider the maximal wrapping, again corresponding to vacua at the root of the baryonic branch of the $N=2$ vacua.  In this case the integral of $C_2$ over the wrapped sphere at any moment in time increases from $0$ when the NS5-brane was first created to $m$ at the end, at which point the volume of the $S^2$ shrinks to zero and so $C_2$ must blow up.  However $C_2$ is not gauge invariant, rather $F+C_2$ is gauge invariant.  This means that we may keep $C_2$ finite if we perform a gauge transformation during the NS5-brane's lifetime that decreases the integral of $C_2$ by $m$ units but increases the integral of $F$ by the same $m$ units.  The integral of $F$ is the worldvolume D3-brane charge, and so when the NS5-brane disappears it carries $m$ units of D3-brane charge.  Of course this choice does just correspond to a gauge choice, which is consistent with the S-dual twisted K-theory fact that D3-brane charge in this background is only conserved modulo $m$.  However the choice of a gauge in which $C_2$ does not blow up appears to coincide with a choice of gauge group such that the gauge coupling is finite.

This is not exactly the process that we are looking for.  The problem is that the dielectric NS5-brane that wrapped the 3-cycle creates a single unit of $H$ flux, that by a choice of boundary conditions we may consider to extend along the 2-cycle and time.  Thus the integral of $B$ over the two-cycle shifts by one unit for each such nucleation as we allow the system to time-evolve.  This is similar to what we found above using T-duality, but the sign is wrong.  The shift \textit{increases} the D3-brane charge in the D5 stack by $m$ units.  The fact that the total D3-brane charge is conserved is S-dual to the following fact.   On a $spin^c$ manifold, if the $H$ flux is sourced by a stack of NS5-branes then a stack of D$p$-branes that is nontrivial in homology but trivial in twisted K-theory may decay, but this decay leads to a worldvolume flux on the NS5-brane stack that carries precisely the lost charge, and so the D$p$-brane charge is conserved whenever the $H$ flux is sourced \cite{Nov}.  In our case the $G_3$ flux is sourced by D5-branes and so D3-brane charge is conserved when one counts the contribution from the MMS instanton NS5-brane and also from the D3-brane charge of the D5-branes.

\begin{figure}[ht]
  \centering \includegraphics[width=5in]{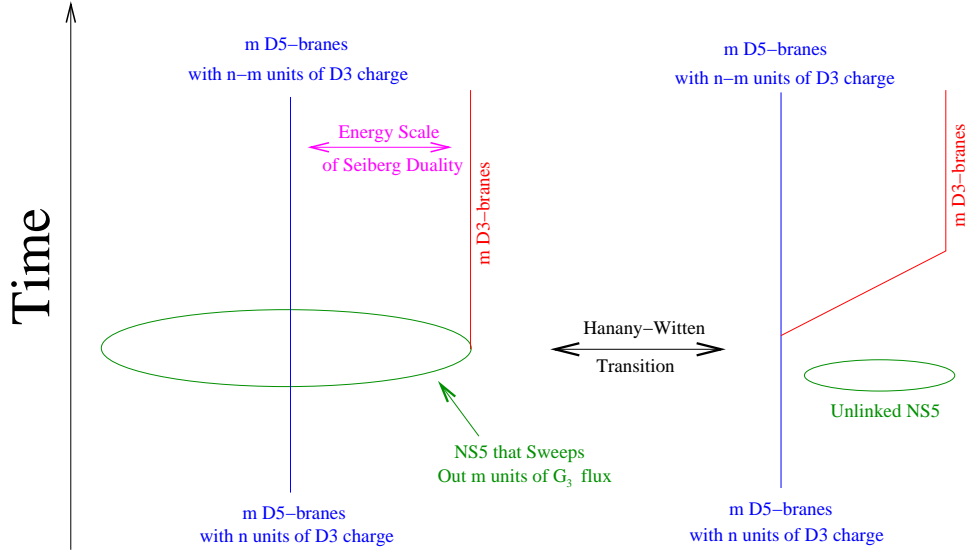}
\caption{The cascade occurs when an NS5-brane bubble spontaneously forms in the vacuum and sweeps out a 3-surface linking the D5-branes.  The NS5-brane wraps a trivial cycle and so carries no net NS5-brane charge, but the $G_3$ flux of the D5-branes produces D3-brane charge on the NS5-branes.  Similarly the $H$ flux of the NS5-branes destroys an equal amount of D3-charge on the D5-branes, reducing the gauge group as seen by a probe not big enough to extend to the NS5-branes from the D5-branes, where the gauge theory lives (the horizon at large $m$).  The fact that the total D3-charge is conserved, as is clear to a UV observer who notices no reduction in gauge group, is apparent from the Hanany-Witten transition to a stack of D3-branes escaping.} \label{cascade}
\end{figure}

To support the claim that the D3-brane charge is conserved we consider the case in which the D3-brane emitted from the NS5-brane is then absorbed by the stack of D5's.  This is related by a Hanany-Witten transition to a configuration in which the D5-branes and NS5-brane do not link at all, in which case clearly no charges could have changed.  In Fig.~\ref{cascade} we see that even the case in which the D3 is not reabsorbed is equivalent to one in which the stack of D3's escape from the D5's, a process which surely preserves the D3-charge.  At finite $g_s$ each D3-brane is then seen to be just the thin tube of dielectric D5 and NS5-brane \cite{bion} created when a D5 and NS5 appear to cross that ensures that in fact they do not cross.  The existence of such tubes ensures that at finite $g_s$ the linking of numbers of NS5-branes and D-branes never change, and so the non-linked situation is continuously related to our situation in which the branes appear to link but are actually unlinked by a stack of D3-branes.

The solution \cite{KPV} is to consider an instantontic NS5-brane with the opposite orientation to that considered above.  This NS5-brane will gain $m$ units of D3-brane charge while the D5-branes will, as desired, lose $m$ units of D3-brane charge.  Again the D3-brane charge is conserved, and so one might think that the rank of the gauge group is not reduced.  However the gauge theory lives on the D5 and D3-branes, and so is coupled even at low energies to the D3-branes created on the D5-brane worldvolume.  However the IR degrees of freedom do not couple the gauge theory to the worldvolume $SU(m)$ gauge theory of the D3-branes contained in the NS5-brane, because the distance (in the $w$-plane, where the dual quarks live) to the NS5-brane is the length scale at which this step of the cascade occurred.  Thus a probe at a higher energy than the energy at which this Seiberg duality occurred would couple to all of the D3-branes and thus not see a reduced rank, but this is what we expect as only probes at energy scales below that of the duality see the reduced gauge group.  

The fact that the distance to the NS5-brane is the energy scale of the Seiberg duality is visible from the T-duality to MQCD.  These NS5-branes are a remnant of the part of the M-theory configuration that we omitted when we reduced our configuration to type IIA.  This is particularly clear if we use the viewpoint in which the RG flows with the radius.  Now the NS5-branes are static and the whole configuration is T-dual to the following improved IIA reduction of the M-theory configuration.  Instead of making the NS5-brane continue, completely flat, forever along the $v$ plane we instead approximate the M5 by a series of steps.  In particular the flat sheet at fixed $x^6$ only continues up to the radius of the Seiberg-duality transition.  At the boundary of this disc we may bend the NS5-brane at a 90-degree angle, and attach a cylindrical NS5-brane which wraps the $x^6$ circle once and then bends again to sweep out an annulus in the $v$ plane, and then another cylinder, and so on approximating the M-theory curve.  Each such cylinder contains $m$ units of D4-brane charge, as it wraps the M-theory circle $m$ times each time that one travels around the origin of the $v$ plane once.  As the cylinder is an NS5-brane that lies along the direction that is T-dualized, its T-dual will be another NS5-brane.  The $m$ units of D4-brane charge carried by the NS5-brane will dualize to $m$ units of D3-brane charge carried by the NS5-brane in IIB.  Intuitively the smearing of these D3-branes along the $w$-plane at the length scale of the Seiberg duality may be responsible for the mass gap of their worldvolume theory, this is in accord with the fact that the gauge symmetry is restored if the $N=2$ supersymmetry is restored by projecting to $w=0$.  

The cylinder was always topologically trivial in IIA, that is it carried no net NS5-brane charge, and so the T-dual will also carry no NS5-brane charge.  Thus we have found our stack of $m$ D3-branes that have dielectrically expanded into an NS5-brane.  As desired the distance to the branes on which the gauge theory lives is the scale of the Seiberg duality because the step was placed so as to approximate the M-theory curve. 

At energy scales above that of the Seiberg duality, probes are larger than the sphere swept out by the NS5-brane.  Thus the probes are larger than the distance between the D3-branes that have disappeared and those that have appeared, and so they do not register any loss of D3-brane charge.  Correspondingly, at energy scales above the transition the rank of the gauge group is not reduced.  



In more general cascades there will be several stacks of D5-branes wrapping the 2-spheres that resolve various singularities.  These stacks will each carry some D3-brane charge, and if the D3-brane charge is high enough then NS5-branes will nucleate and then self-destruct, sweeping out a contractible 3-cycle that links some subset of the stack.  Each linked D5-brane will lose a unit of D3-brane charge, and this will continue until such decay processes are no longer energetically favorable.

At large $n$ the D5-branes are at the horizon.  However the $G_3$ flux created by the D5-branes remains and the integral of the $G_3$ flux over the 3-cycle is still equal to $m$.  Thus an NS5-brane sweeping out the 3-cycle still must emit $m$ D3-branes before shrinking out of existence.  Now the D3 charge of the D5's at the horizon drops and so the rank of each simple group has been reduced by $m$ units.  However at large $n$ it is no longer clear how an NS5-brane may link less than the maximal $m$ units of flux, and so the nonbaryonic roots are obscured, if indeed such vacua survive the large $n$ limit. 

\section{Conclusion}

We have argued that the WZW-like proposal for a continuous description of the cascade using the dielectric effect, in which the D3-branes embedded in the D5-branes decay via NS5-brane instantons, does not decrease the rank of the gauge group as desired.  Instead it creates a flux that increases the rank of the gauge group in the IR, although at the original energy scale the gauge group is unaffected.  Instead we have seen that the inverse process, in which dielectric NS5-branes nucleate and yield D3-branes, creates a flux that leads to less D3-brane charge embedded in the D5-branes where the gauge theory lives.  Thus probes that are not sufficiently energetic to reach the NS5-branes will detect only the decreased D3-brane charge in the D5-branes, and will see the rank of the gauge group reduced as desired.  The KPV proposal has the drawback that each step of the decay is a transition from a nonsupersymmetric configuration with anti-branes to a supersymmetric configuration, and so it may only proceed one step.  In fact its range of validity, $m>>m-n$, ensures that it may only proceed for one step.  Here no such restriction exists as the driving force for the process does not consist of anti-branes which are inserted by hand, but rather the tension of the M5-brane which is visible in the M-theory (and probably F-theory) lift but not in the IIA or IIB configuration.  However by T-dualizing snapshots of the M-theory configuration reduced to IIA at different energy scales the effects of the truncated M-theory dynamics are retained.

Our argument is supported away from the large $N$ limit, where traditionally such discussions have taken place, because it relies only upon the T-duality to MQCD.  While the T-duality to MQCD has been heavily exploited in large $N$ approaches \cite{KS,OT}, it is unclear to what extent the MQCD approach is valid at large $N$.  One obstruction is that the usual argument that allows one to pretend that spacetime is flat by introducing harmonic coordinates \cite{Gibbons} needs to be consistent with the large $N$ geometry.  The large $N$ limit has been critical to previous approaches, being used for example to justify the deformation of the conifold.  However at radii smaller than the smallest $|v|$ of the curved M5-brane there can be no NS5-branes in the IIA reduction and so the T-dual does not contain a $A_1$ singularity.  Thus one might suspect that the T-dual at such small radii, which correspond to energy scales below the last Seiberg duality, is a smooth space, and in particular that the conifold singularity is not there.  Thus perhaps the deformation of the conifold is a straightforward consequence of the MQCD configuration.  Of course such a minimum $|v|$ only exists if we turn on a superpotential, but this is when we expect the chiral symmetry breaking caused by this deformation.

In addition, as this argument is largely topological, it extends to $N=2$ theories broken to $N=1$ by an adjoint chiral multiplet mass $\mu$.  The original case is recovered in the limit of large $\mu$.  These configurations, in turn, are local descriptions of more general theories in which the $N=2$ supersymmetry is broken to $N=1$ by a superpotential which is polynomial in the adjoint chiral multiplets and sufficiently nondegenerate.  Thus the T-duality described above may be applied to a neighborhood of the extrema of any such superpotential, the only places where nonconformal products of two simple gauge groups with bifundamental hypermultiplets may be found.  The algorithm generalizing this approach to more than two gauge groups at the same extremum has been described above. 

We have also seen that in the MQCD picture, in addition to the baryonic root vacua (with the $r=0$ vacuum chosen at the last step) that define the Klebanov-Strassler cascade, there are many other cascades corresponding to different vacua.  In particular we have described the simplest cascade,
\begin{equation}
SU(n)\times SU(n+m) \longrightarrow SU(n)
\end{equation}
which corresponds to the $r=0$ nonbaryonic root of the $SU(n+m)$.  Combining this with the baryonic root description, we can conclude for example that the Klebanov-Strassler cascade can stop at any step if the $r=0$ vacuum is chosen.  It may be interesting to use the MQCD picture to classify all of the cascades and to compare them with the field theory, where one has a choice of vacua after each duality.

An independent application is to the twisted K-theory classification of D-branes, where it is a first example of three phenomena.  First, this example is evidence for an S-duality covariant generalization of the K-theory classification, and in fact is inconsistent with the usual K-theory classification.  Second, the fluxes produced by the MMS instanton, which are sometimes dismissed as irrelevant,  here are seen to play a key role.  The lost D3-brane charge on the D5-stack in fact is not destroyed directly by the MMS instanton, but on the contrary is destroyed by the flux created by the instanton.  In fact the cancellation between the charge-violation of an MMS instanton and the charge violation on the worldvolume of the source of the flux it wrapped plays the physical role of ensuring that in the UV the original gauge group is restored.  That is, no D3-branes are really ever destroyed, it is just that our probes get shorter and so we can not reach them all anymore as we flow into the IR.

Finally in the large $N$ near-horizon case we see the (S-dual) K-theory classification at work, classifying fluxes.  However the branes classified by the (S-dual) K-theory are not in the spacetime at all, but rather live at the horizon.  Previously K-theory classifications of D-branes were restricted to finite stacks that live inside the spacetime, in fact in the approach of \cite{MW} it is crucial that the branes are inside of some sphere.  

The two big questions are left untouched.  First, what is the S-duality covariant K-theory classification?  Second, what is the F-theory lift of the conifold configuration?  It is easy enough to fiber the torus over 9 of the dimensions, this just gives the M-theory configuration.  But is its fibering over the IIB circle even well-defined?  The IIB circle consists of the set of maps from the F-theory torus to the $E_8$ fibered over M-theory, so a point in the total space consists of a 9-dimensional point plus a point in the $E_8$ fiber and so a section of an $E_8$ bundle over the 9d space with the F-theory branes describing the obstructions to creating this bundle from the 11d bundle.  But when is such a construction possible?



\bigskip\bigskip

\noindent 
{\bf Acknowledgements}

\noindent  
I would like to thank Nick Halmagyi for posing this problem, and Sunil Mukhi for telling me that KPV solved it first.  I would also like to thank M. Douglas and I. Klebanov for discussing this proposal.  This work has been partially supported by IISN - Belgium (convention 4.4505.86), by a ``P\^{o}le d'Attraction Universitaire'' and by the European Commission RTN program HPRN-CT-00131, in which I am associated to K. U. Leuven.

\noindent


\end{document}